# A metapopulation model of the spread of the Devil Facial Tumour Disease predicts the long term collapse of its host but not its extinction

## Authors

- Veronika Siska[1]
- Anders Eriksson[1,2]
- Bernhard Mehlig[3]
- Andrea Manica[1]

## Affiliations

1. Department of Zoology, University of Cambridge, Downing Street, Cambridge CB23EJ, United Kingdom
2. Department of Medical & Molecular Genetics, King's College London, Guys Hospital, London, United Kingdom
3. Department of Physics, University of Gothenburg, SE-41296 Gothenburg, Sweden

## Abstract

The Devil Facial Tumour Disease (DFTD), a unique case of a transmissible cancer, had a devastating effect on its host, the Tasmanian Devil. Current estimates of its density are at roughly 20% of the pre-disease state, and single-population epidemiological models have predicted the likely extinction of the host. Here we take advantage of extensive surveys across Tasmania providing data on the spatial and temporal spread of DFTD, and investigate the dynamics of this host-pathogen system using a spatial metapopulation model. We first confirm a most likely origin of DFTD in the north-east corner of the island, and then use the inferred dynamics to predict the fate of the species. We find that our medium-term predictions match additional data not used for fitting, and that on the long-term, Tasmanian Devils are predicted to coexist with the tumour. The key process allowing persistence is the repeated reinvasion of extinct patches from neighbouring areas where the disease has flared up and died out, resulting in a dynamic equilibrium with different levels of spatial heterogeneity. However, this dynamic equilibrium is predicted to keep this apex predator at about 9 % of its original density, with possible dramatic effects on the Tasmanian ecosystem.

# Introduction

There is growing recognition that population structure and spatial heterogeneities can play a very important role in determining the transmission and persistence of pathogens[1]. Whilst spatially-explicit models of human diseases are becoming increasingly common, similar approaches for wildlife diseases are more limited as it is rarely possible to obtain the large amount of temporally and spatially tagged data necessary to fit such models. The Devil Facial Tumour Disease (DFTD) offers an ideal study system to investigate the importance of the spatial component in the epidemiology of a wildlife disease.

DFTD is a transmissible cancer that affects the Tasmanian Devil[2], and is spread through biting during social agonistic interactions[3]. This cancer was first detected in 1996, and it has since had a devastating effect on its host[4], with host density decreasing by approximately 80% across the island and local populations crashing to less than 10% of pre-disease levels[5]. Whilst no local extinction has been recorded yet, a number of single-population epidemiological models have predicted the likely imminent extinction of Tasmanian Devils as a result of DFTD[4,6,7], due to the frequency-dependent nature of transmission, where the number of contacts does not decrease with decreasing population size[4]. Although there are signals of rapid genetic adaptation[5] and there have been cases of devils mounting an immune response[8], in the vast majority of the cases, the cancer leads to death[9].

The DFTD's recent emergence has allowed comprehensive recording of the decline of Tasmanian Devils in response to the disease. Extensive data on the spatial spread of DFTD since 1996[10], together with detailed information on the local host-disease dynamics from a number of locations[4,9,11] (Fig. 1), offer the possibility of investigating the metapopulation dynamics of this system. The key question is whether there is a rescue effect[12,13] strong enough to prevent the extinction of the host in a system where all single-population models predict its demise within a relatively short time frame.

# Results

Here we present a spatially-explicit stochastic metapopulation model representing the whole island of Tasmania (Supp. Fig. S1.). We used detailed maps of pre-disease host density to inform the dynamics of local populations[10], and data from the literature covering the period from the first detection of DFTD up to 2007 to inform a number of demographic parameters[14] (Table S2). Because juveniles and subadults are known to suffer very low infection rates compared to adults (possibly due to their limited participation in fights)[4], we included host age as a factor in the local dynamics of each population, similar to previous modelling work[14]. The local demography was governed by logistic growth and the disease-spread by a compartmental model with frequency-dependent transition rates, where infection leads to certain death through an exposed, a diagnosable (but not yet infectious) and finally an infectious state. These local populations were coupled by migration (permanent change in the home-range) and contact (infection between neighbouring patches, representing overlapping home-ranges), both of which were essential for the combined population dynamics of the host and the disease.

We used an Approximate Bayesian Computation framework[15] to find the model parameters that best fitted the data on the spatial spread of DFTD and its local temporal dynamics (Fig. 1), pinpointed its likely geographic origin, and estimated unknown disease-related parameters (lifetime of compartments and infection, migration and contact rates). Finally, we used the fitted model to study the possible long-term persistence of both the host and the disease.

Our model captured the speed of the spatial spread of the disease across the island (Fig. 2 A & B), as well as the local dynamics in terms of both the rate of increase in disease prevalence and its equilibrium frequency (Fig. 2C). The reconstructed spatial spread of the disease provided a reasonable qualitative fit with available data (Fig. 2A), and confirmed the geographical origin of the disease in the north-eastern corner of Tasmania, in the area where the first diseased animal was observed[10] (Fig. 2D). An aspect of the spatial spread that our model failed to match is the late appearance of DFTD in the north-western corner of Tasmania[9] (Fig. 2A). However, this mismatch was to be expected, since this delay has been ascribed to two factors not present in out model: spatial barriers and the initial dominance of the tetraploid karyotype in the area, which was associated with lower virulence compared to the more common diploid karyoype[16]. The latent period was within the plausible range from the literature: a median of 7.7 months with [1.2-11.3] 95% CI versus 3-9 months[4,14] (Supp. Fig. S6). On the other hand, the diagnosable period was longer than previously reported: a median of 10.9 months with [5.6 -14.2] CI (mean with 95% confidence interval) vs less than 3 months[4,14] (Supp. Fig. S6). However, we note that previous estimates of the diagnosable period were speculative or based on infrequent (at most quarterly) sampling[4].

We validated our model by using data from 2008 till 2016 collected from a number of sites, including five new locations not used for fitting[17]. We examined five independent re-runs from the best-fitting 200 parameter sets: medium-term predictions for the decline of population densities were in agreement with the latest observed field data[17], both for local populations (Fig. 3A) and for the whole island (Fig. 3B). Most observations are within the 95% confidence interval of the model prediction and support a long-term population decrease. Since field data from population densities were not used during the fitting procedure and this comparison even included a temporal extrapolation (2007 to 2016), this test highlights the predictive power of our model.

We further tested the long term dynamics by looking at the longer term predictions of the five independent re-runs from the best-fitting 200 parameter sets, for which we collected data for a total of 250 years. Strikingly, all of the 200 best-fitting scenarios predicted the long term coexistence of host and pathogen. The long-term population size, after the system reached equilibrium (~75 years after the initial mutation, Supp. Fig. S7-S8.), was reduced to a median of 9% of the original, pre-DFTD size, with a median prevalence of 41 % (Fig. 4A & B). Furthermore, the resulting age-structure was highly distorted by a shift to the younger age classes, qualitatively similar to what has been observed in the field[11,18] (Fig. 4C and Supp. Fig. S9.).

Inspection of the scenarios supporting long-term coexistence revealed a broad range of large-scale spatial and temporal dynamics, ranging from relatively stable prevalence (Fig. 5A) with a homogeneous spatial distribution (Fig. 5B) characterised by relatively stable local dynamics (Fig. 5C), to strong cycles (Fig. 5D) and a highly heterogeneous spatial distribution of the disease (Fig. 5E) with highly unstable local dynamics (Fig. 5F). The latter was accompanied by local extinction and recolonization events, where migration enabled the recolonization of areas wiped out by the cancer, leading to the persistence of the overall population through the so-called metapopulation rescue effect. The magnitude of cycles was highly variable between different parameter sets, and was largely determined by quantities linked to the local disease time-scale, especially the latent and infectious periods and the magnitude of the migration rate (Supp. Fig. S10.). Short disease-related life expectancy and weak linkage between patches through migration lead to higher spatial heterogeneities and stronger temporal fluctuations. This was due to the quick, local burnouts of the devils by the cancer destabilising the system, and a strong spatial autocorrelation of the cycles that created large interconnected clusters of populations going extinct together. For shorter diagnosable periods, as observed in the field, this means a higher likelihood of unstable dynamics with large cycles.

## Discussion

Our aim was to create a spatially explicit model to reproduce the main characteristics of a disease spreading in a spatially structured population and use it to make long-term predictions. Our model produced a good fit to the data and medium-term predictions in agreement with data that were not used during fitting. Our model contradicted the long-term predictions of previous models describing local populations: such models predicted the extinction of the host, but the spatial dimension changed the outcome so that the host and the pathogen co-exist in a dynamic equilibrium. This persistence of both host and pathogen was due to the metapopulation rescue effect, in which migrants from coupled populations lower the risk of extinction both locally and for the overall population. Even when local populations go extinct, they can be recolonised by healthy individuals from neighbouring populations, ensuring the persistence of the metapopulation as a whole. The observed slight decrease in population density at one of the sites (Fentonbury) 8-10 years after the arrival of DFTD[17] could be the sign of such a fluctuation. This radical change in the long-term predictions for a real-world system highlights the importance of spatially explicit models in wildlife epidemiology.

Our model was a simplified representation of DFTD: both the pathogen and the hosts were all identical, and there were no seasonal effects or spatial barriers. We considered a single unchanging cancer strain, but cancers are known to have a complex evolutionary dynamics, with multiple strains[16], and even multiple different cancers[19]. The behaviour of hosts was also assumed to be identical and constant in time and did not account for the variation in host immunity[5,8], the increased effect of DFTD on individuals with a higher fitness[20], changes in life-history (e.g. the observed earlier maturation[18]), differences between the sexes in social behaviour (e.g. males are known to migrate further than females[4]) or the social network of devils (although, based on field data, sex was shown not to have a significant effect on prevalence[4] and the social network was close to a random graph[21]). Both demography and disease spread are known to have some seasonal components[4] which we did not take into account. However, the seasonal component in prevalence was shown to diminish with age and disappear for mature animals over 2 years, most likely meaning that the effect is from the seasonal replenishment of susceptibles and not from a seasonal forcing of infection[4]. Finally, spatial barriers likely play a role in the late arrival of DFTD to the north-west corner of the island[16], which is not captured by our model.

In order to address the limitations of this simple spatial model, more data are needed, particularly denser spatial sampling across Tasmania so that spatial inhomogeneities can be accounted for, as well as strain-specific information to enable more accurate models of the cancer. With such data, it will be possible to evaluate whether natural selection is acting, or if stochasticity is enough to explain the observed rich dynamics of various cancer strains. Furthermore, the model could be used to evaluate possible control strategies, such as culling, vaccination, releasing healthy animals from insurance populations or installing artificial spatial barriers.

In this study, we created a spatially explicit metapopulation model for the spread of DFTD, which reproduced the observed rapid dynamics of DFTD and its devastating effect on the population of Tasmanian devils. In contrast with local models, it predicted the long-term persistence of both the cancer and its host, but the likely scenarios all imply very low population levels of the Tasmanian devil (median of 9% of the original, healthy population size). Such a decline in population size of an apex predator can have the same severe destabilising effect on the ecosystem as their extinction[22]. The large fluctuations in population size and the distorted age-structure could compound these problems through the loss of genetic diversity and rise in the frequency of harmful genetic variants in the populations and by magnifying the effects of other threats like habitat loss, predation by feral cats and roadkill, and thus point to the necessity of further efforts in the conservation of this unique species.

# Figures

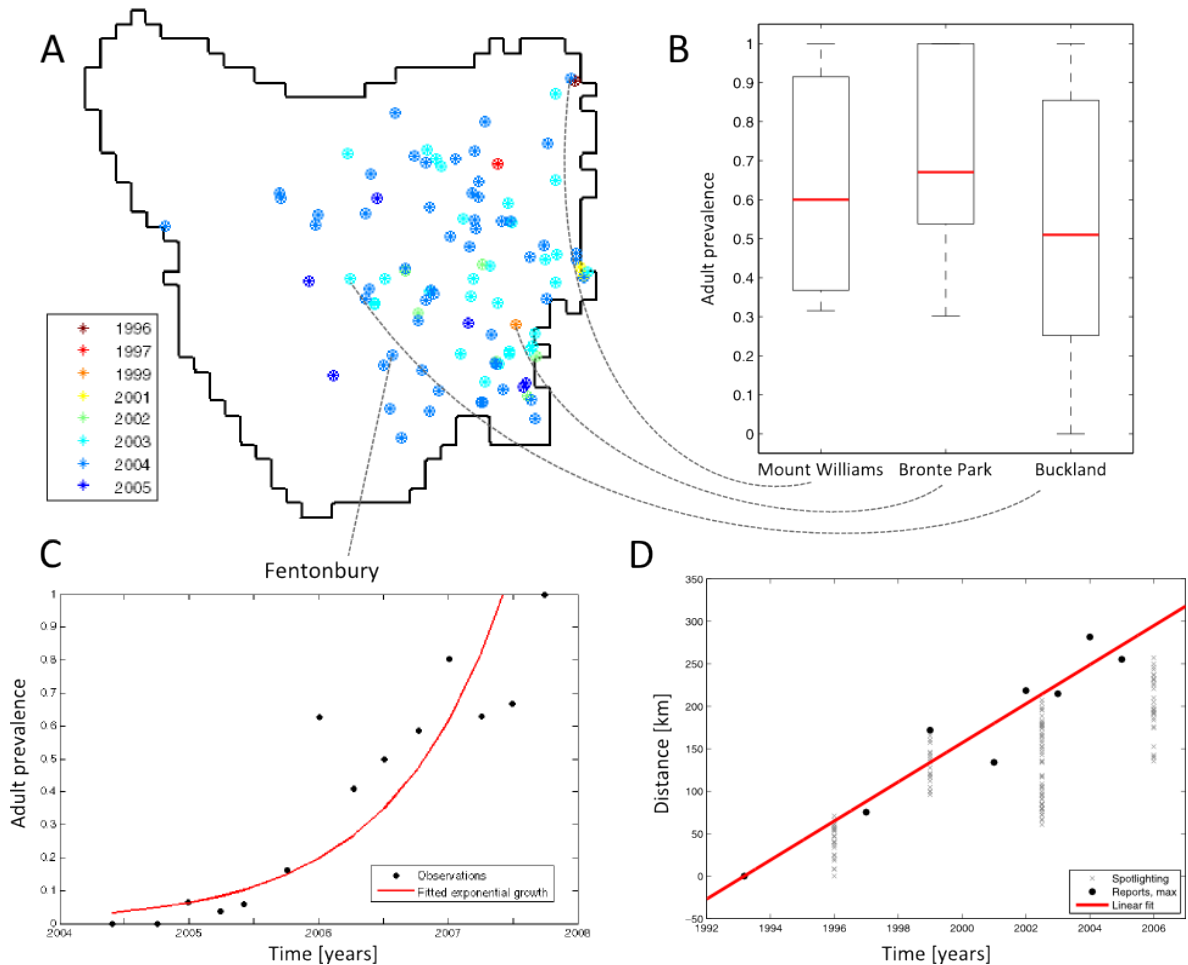

**Fig.1. Observed spread of the disease.** Based on data from McCallum et al.[4]. A. Reports of diseased animals in Tasmania, color-coded according to the year of the report. Also showing sites used to estimate summary statistics. B. Data from Buckland, Bronte Park and Mount Williams, used to estimate the local equilibrium prevalence (0.6424). Red line marks the median, black boxes span from the 25th to the 75th percentile and whiskers extend to the most extreme data points. C. Data from Fentonbury, used to estimate the local initial growth rate in 2-3 year-olds (2.2644 1/y). Black dots mark observations from the field data and red line the fitted exponential growth. D. Fitted spatial speed of the spread, using the location of the first report as the origin. Gray crosses mark data from annual spotlighting survey, not used for the fitting. Black dots show the distance to the furthest report in that year and red line the result of a linear fit to these points. In the simulations, the fitting was redone for each spatial origin, resulting in a different intercept (time of the first mutation) and slope (speed of the spatial spread) for each origin.

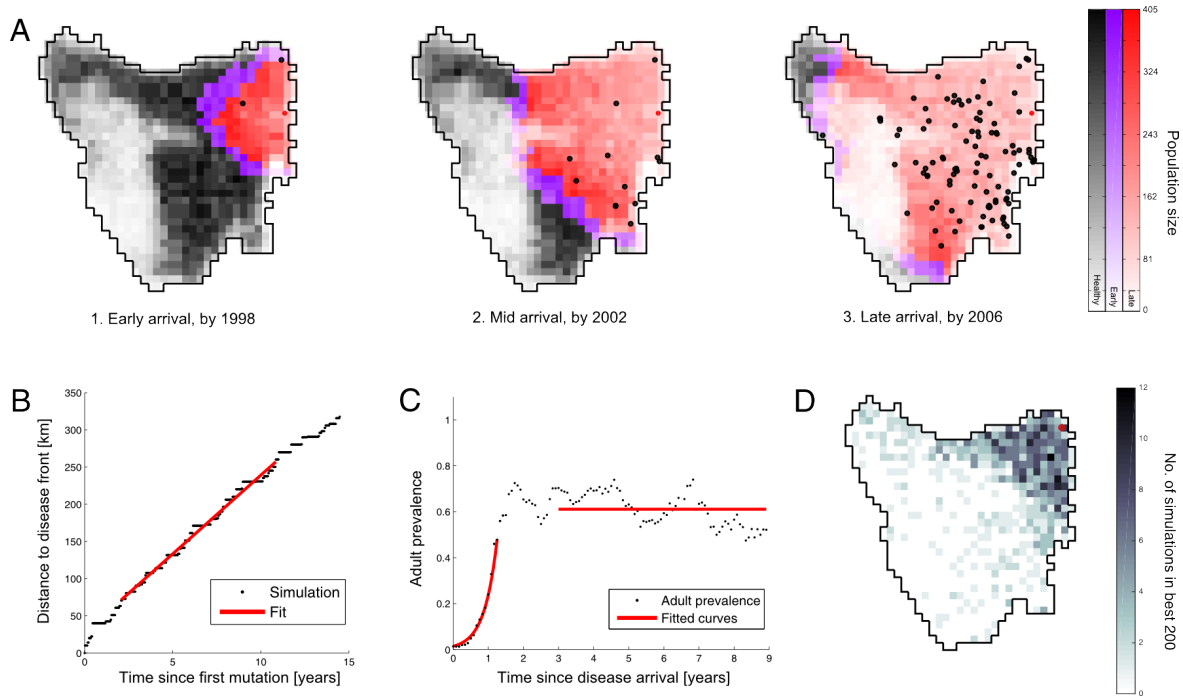

**Fig. 2. Simulated spread of the disease from the best-fitting parameter set**. A. Map showing snapshot of the epidemic, overlaid with reports of diseased animals up to that year (black dots) and the origin of the disease (red dot). Population spread is indicated by the shade, while healthy (black), early-phase (purple) and late-phase (red) patches are color-coded. Healthy populations are those without any diseased animals, while populations with at most 40% prevalence and a population size of at least 50% of the local carrying capacity were classified as early phase. B. Distance from the origin to the disease front and the fitted speed of the spatial spread. C. Local time series at Fentonbury and the fitted initial increase rate in prevalence and equilibrium prevalence. D. Distribution of the spatial origin of the 1000 best-fitting parameter sets.

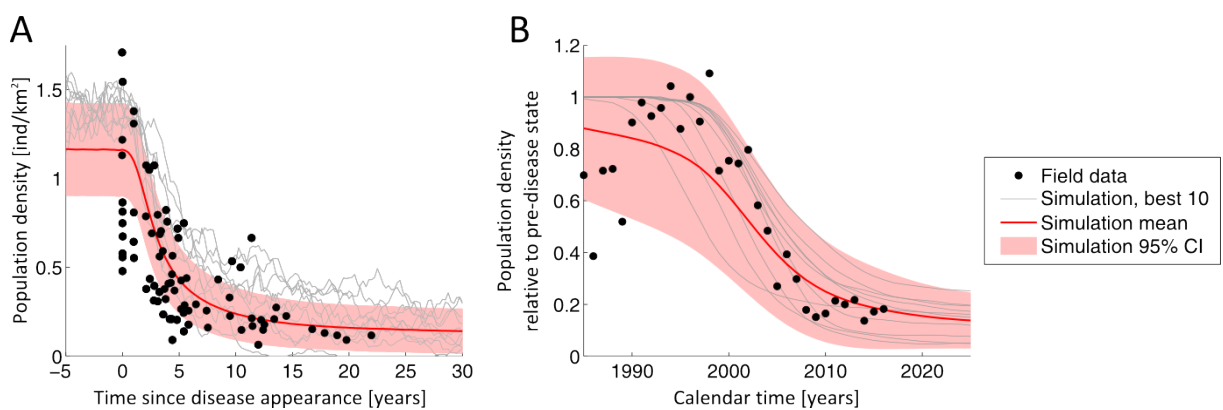

**Fig. 3. Testing medium-term predictions for population density.** Field data (black dots), which were not used to fit the model, overlaid with time-series from a single run of the 10 best-fitting parameter sets (gray lines) and the mean (red line) and 95% confidence interval (red shaded area) from 5 re-runs of the 200 best-fitting parameter sets. A. Data from trapping surveys and simulation data from local populations at the same location, in quarterly bins. B. Data from spotlighting surveys and simulation data from the whole island, in yearly bins.

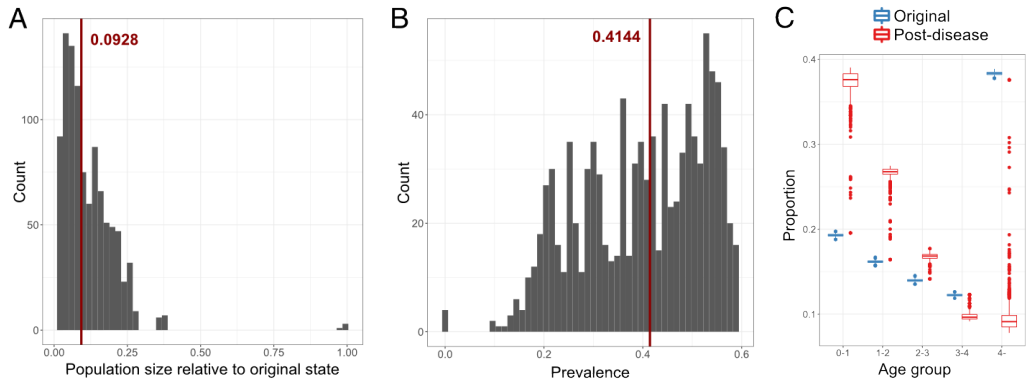

**Fig. 4. Long-term predictions**. Results from 5 independent runs of the 200 best simulations. A. Histogram of the population size relative to the original state of 60,000 animals, with the median indicated in red. B. Histogram of adult DFTD prevalence, with the median indicated in red. C. Average age-structure of the total population from the pre-disease and long-term diseased state.

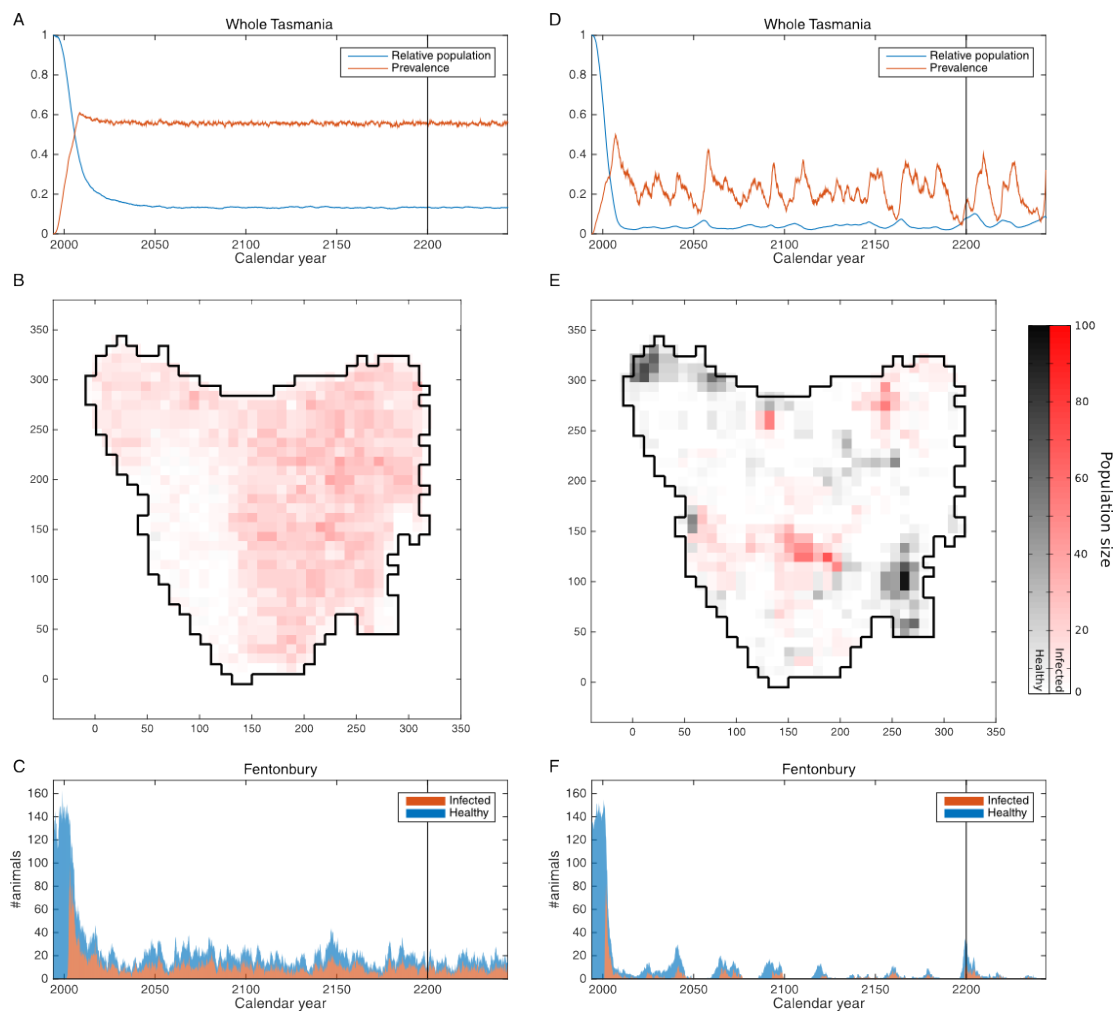

**Fig. 5. Simulated long-term dynamics**. The best fitting stable (A-C) and strongly heterogeneous (relative standard deviation of prevalence over 20%, D-F) cases. A, D. Time-series of relative population and prevalence from the whole island. B, E. Time-series of the number of infected and healthy devils in Fentonbury. C, F. Snapshots from the simulation. Population spread is indicated by the shade, while healthy (black) and infected (red) patches are color-coded.

# Methods

## Field data

### Demographic parameters
To set the local demography, we used the birth and death rates in an age-structured model with five yearly age classes, as estimated from current knowledge of devil life history and some modelling by Beeton & McCallum[14] (Supplementary Table S1.)

We also used published data to set some spatial characteristics of the model. We used the probability of occurrence of Tasmanian devils in 1 km$^2$ blocks, presented by Hawkins et al. [10], as a proxy for the pre-disease population density. This density map was predicted by species environmental domain analysis on the basis of results from a combination of presence survey techniques, and adapted from Jones & Rose[23]. We then scaled these densities such that the overall pre-disease population size of devils in Tasmania was 60,000[24].

Last, to obtain estimates of the ratio of disease transmission rates between different age classes, we used published scaling factors from McCallum et al.[4] as used by Beeton & McCallum[14]. Animals below 1 year-old were not susceptible or infectious, the rate for those in the 1-2 year-old age class were scaled by a factor of 0.602, while the rest of the age classes had the same rate.

### Summary statistics
We also used already published field data to fit our model. We based the local dynamics on population and prevalence time-series from McCallum et. al.[4], covering the period up to mid-2007. For the initial increase rate in prevalence, $r_0$, we only used the published value of $r_0$=2.2644 estimated from 2-3-year old devils in Fentonbury, as the most reliable site and age group and also the estimation used in previous modelling to investigate the effect of culling[14]. Most sites were not monitored early enough, but even other sites where the early disease dynamics were observed violated basic assumptions of our model: they had internal spatial structure (Freycinet Penninsula), were subject to a control trial of culling (Forestier) or had an unusual behaviour, likely owing to strain dynamics (West Pencil Pine). To obtain an estimate of observed equilibrium prevalence, we used the mean from all observations and sites where infection appeared long before the start of monitoring from the same publication (Bronte Park, Buckland and Mt William), resulting in $p_{eq}$=0.6424.

Finally, we calculated the observed linear speed of disease spread using the dates and locations of reported diseased animals from Hawkins at al.[10]. Since this speed depended on the geographic location of the original mutation, which was a free parameter in our model, we performed the following procedure. We first chose a range of possible times: from 1800 to 2000, in monthly increments. Then, for each possible temporal and spatial origin, we

1. Created a time-series of the distance to the report furthest from the origin
2. Performed a least-squares linear fit
3. Calculated measures of goodness of fit (proportion of variance explained ($R^2$), Pearson and Spearman correlations)

Finally, we chose the starting time that resulted in the R2 for each spatial origin and recorded $R^2$, the time of the original mutation and the speed of the spatial spread. Supplementary Figure S3. shows

$R^2$ for each cell, Supplementary Figure S4. the estimated speed of the spatial spread and Supplementary Figure S5. the estimated time of the original mutation for the starting time with the highest $R^2$ value. Although locations in the north-west did not have the highest $R^2$, they had the most recent starting times and hence higher speeds, which were more compatible with the observed quick local spread, hence resulting in better-fitting parameter sets (Figure 2).

We decided to use data from the spotlighting survey in the same publication only as an independent check, since the reports and the spotlighting survey (which was an estimation of disease arrival based on spotting a sharp decline in the estimated population size) showed qualitatively equivalent disease arrival dates, but the latter was less detailed on the temporal scale and was only an indirect measurement of disease arrival (hence less reliable).

**Testing medium-term predictions**
To test our model, we compared its medium-term predictions against field data not used for fitting, from 2007 to 2016, as published by Lazenby et al.[17]. We compared estimates of the population density, an observable not used for fitting. We used both data sources presented in Lazenby et al.[17]: the statewide spotlighting survey, where animals are counted along country roads, and capture-recapture data from trapping surveys at nine sites across Tasmania. By the time of the survey, DFTD has still not reached one site, Woolnorth on the north-western edge of Tasmania, so we excluded that from further analysis. The sites used partially overlapped with those used for fitting (Fentonbury, Buckland and Bronte Park), but also included additional sites (Takone, Granville, Narawntapu, Wukalina and Kempton). Data from the spotlighting survey were recorded in terms of the number of sightings per 10km, which we converted to counts relative to the pre-disease state to be comparable with our model. We took the estimate at the time of the first reported case of DFTD (2006) as the baseline count because our simulations started in that time period (the exact conversion between simulation and calendar time depends on the origin).

# Metapopulation model
Our spatially explicit model (Supplementary Figure S1) consisted of an individual-based (thus stochastic) metapopulation of local populations, arranged on a square grid of 10km by 10km cells representing mainland Tasmania. The local populations on each grid were assumed to be well-mixed and only nearest-neighbour interactions (contact and migration) in the von Neumann neighbourhood (4 immediately neighbouring cells) were allowed. The carrying capacity of each cell was set using published data, as described below.

The local dynamics consisted of a compartmental epidemiological model built on demography following logistic growth. There were four disease-related compartments: susceptible (S), exposed (E, no symptoms or infectivity), diagnosable (D, show symptoms but not infective) and infective. The cancer was assumed to lead to certain death, since very few recoveries have ever been observed[4]. We used a frequency-dependent transmission rate, since it was shown to fit the data better than a density-dependent alternative[4]. All delays (average healthy lifetime, transfer time between age classes and epidemiological compartments) were assumed to be exponentially distributed to make the stochastic simulation computationally efficient. The transition rates are shown in Supplementary Table S2.

All demographic parameters (birth and death rates) were taken from the literature[14] and all disease-related parameters were estimated through the fitting process, except for the scaling matrix of the infection rate between age classes.

The local populations were coupled in two different ways: by migration and contact. In a migration event, individuals permanently moved to a neighbouring patch, representing a change in their home-range. With contact, individuals living on neighbouring populations could infect each other with a rate lower than that within the cell, but stayed at their current population. This corresponded to overlapping home ranges that can cross the (arbitrary) cell borders.

The model was simulated using the continuous-time Gillespie algorithm[25] (Supplementary Figure S2). The algorithm consisted of an initialisation of the state of the system and the rates of each event, and then performing Monte Carlo steps until the predefined ending time of the total simulation was reached. In each step, first the next event was chosen (e.g. a devil on a given cell is infected), with a probability proportional to its rates. Then, a time interval until the event was drawn from an exponential distribution with its parameter equal to the sum of all rates. Finally, both state and time was updated before performing the next Monte Carlo step.

The model was implemented in Matlab version R2015b, which was used both for the fitting procedure and to record the long-term behaviour of the best 200 parameter sets. The code for the model is publicly accessible on Github: https://github.com/siskavera/tasmanian-devil).

## Model fitting

We then fitted the model to the observed data using Approximate Bayesian Computation (ABC), a method to compute the posterior distribution of a model's input parameters given a set of summary statistics with observed values from the data. ABC first takes a prior distribution over all input parameters. Then, input parameter combinations are drawn from this prior, the simulation is run, and summary statistics from the simulation are recorded. For each of these runs, the distance between the observed and simulated summary statistics is calculated, and a pre-defined number of the closest simulations are retained. Finally, the most likely input parameters and the covariance matrix of noise is determined through multiple linear regression between summary statistics and input variables in the retained space (thus assuming a general linear relationship). Finally, the posterior marginal distributions of all input parameters are calculated, given the observed summary statistics and the fitted statistical model.

We used uniform prior distributions for all numerical input parameters (the limits for each parameter are shown in Supplementary Table S3.), and chose the cell of the original mutation with equal probability out of those with a carrying capacity of at least 20, and performed 100,000 independent simulations. Our distance measure for rejection sampling was the Euclidean distance of the normalised simulated summary statistics from those observed. We retained the best 1000 simulations to calculate the posterior distributions, using ABCestimator from ABC toolbox[15] with default parameters.

Our input parameters were the following:

1. Local infection rate: $\eta$

2. Latent period: $\tau_E$
3. Total diagnosable period: $\tau_D+\tau_I$
4. Proportion of infective period to total diagnosable period: $\tau_I/(\tau_D+\tau_I)$
5. Between-patch contact rate: $c$
6. Migration rate: $m$

The simulated summary statistics (see below for details of how we obtained estimates from the field data) were:

1. Early growth rate of prevalence (proportion with symptoms): $r_0$.
2. Equilibrium value of prevalence: $p_{eq}$.
3. Speed of the disease front: $v$

In the simulations, $r_0$ and $p_{eq}$ were calculated from recorded time-series from the cells in which the field sites were located, each with carrying capacities between 150 and 200 animals (Fentonbury: 196, Mt William: 148, Bronte Park: 173 and Buckland: 194). Time-series were recorded from when the number of diseased animals reached five, in order to avoid excessive noise from unsuccessful local outbreaks.

Then, an exponential curve was fitted to the first 1.5 years of the outbreak to determine $r_0$ amongst 2-3-year-old animals, and the average adult prevalence in the period from 3 years to 9 years was calculated for $p_{eq}$. These periods were chosen to work well for both simulations (with parameters giving a good fit) and data, based on manual examination of time-series and considering the time-scales of the parameters from the data ($1/r_0$). For $r_0$, the chosen period captured the initial phase before the exponential growth takes off. For the equilibrium prevalence, we aimed for a period after it reaches a stable level but before the population size drops too low (below ~10% of the original value) leading to stochastic fluctuations due to small numbers. For the equilibrium value of prevalence, we calculated the mean from the union of long-term time-series from the three available sites.

Finally, to determine $v$, the spatial speed of the disease spread, we estimated the slope from the linear fit as a function of time to the distance of the diseased site furthest from the origin. We calculated the total time to reach the maximal spatial extent in the simulation and cropped the first 1/6th and last 1/4th. The former was to disregard the initial transient for the establishment of DFTD and the latter to not take badly connected areas into account, both of which were unobserved in the field data used for fitting the model.

## Medium and long-term behaviour

To investigate the medium and long-term behaviour of our system, we used the best-fitting 200 simulations and performed five independent runs for each, lasting 250 years after the first mutation. We measured the probability of extinction for the cancer and the host and recorded the number of healthy and diagnosable animals in each age class and local population.

To compare medium-term predictions of population sizes from our model to the capture-recapture data from trapping surveys, we calculated the population density for the eight sites considered. This data was then averaged over 3-month periods to match the frequency of trapping surveys, which was up to four times a year. When comparing model output to yearly spotlighting surveys, we calculated the mean total population size across the island in each year. For each simulation, we converted simulation time to calendar time using the estimated calendar time of the first mutation. This was acquired as part of the fitting procedure and depended on the origin of the disease: it was the time that produced the best fit to the speed of the spatial spread of DFTD between field data and simulation output, assuming a given spatial origin.

After examining the overall dynamics, we decided to consider data from 75 years after the initial mutation onward to calculate statistics describing the long-term behaviour. To further investigate the heterogeneity of the long-term state, we calculated the standard deviation and mean of long-term adult prevalence in the overall population. We then used the mean relative standard deviation from the independent runs per set of input parameters as the proxy for the cycle strength. We performed random forest regression with a forest consisting of 400 trees, and calculated and plotted two variable importance measures, using the randomForest 4.6-12 R package[26]. The first measure is the normalised difference in mean squared error on the out-of-bag portion of the data, after permuting the variable. The difference is recorded for each tree and normalised over the forest. For the second measure, the total decrease in node impurities (as measured by the residual sum of squares) from splitting on the variable is recorded for each tree, and averaged over the forest.

## Acknowledgements

**Funding:** V.S. was funded by the Erasmus Mundus Programme and the Gates Cambridge Trust.

**Author contributions:** V.S. performed the modelling and assembled data from the literature, with input from all co-authors. V.S., A.M., A.E. and B.M. wrote the manuscript.

**Competing interests:** The authors declare that they have no competing interests.

**Data and materials availability:** The code for the model is publicly accessible on Github: https://github.com/siskavera/tasmanian-devil).

**Acknowledgements:** The authors would like to thank Hamish McCallum, Elizabeth Murchison and the organisers of the Erasmus Mundus Masters for Complex Systems Science for their advice and helpful comments.

# Supplementary figures and tables

## Metapopulation model

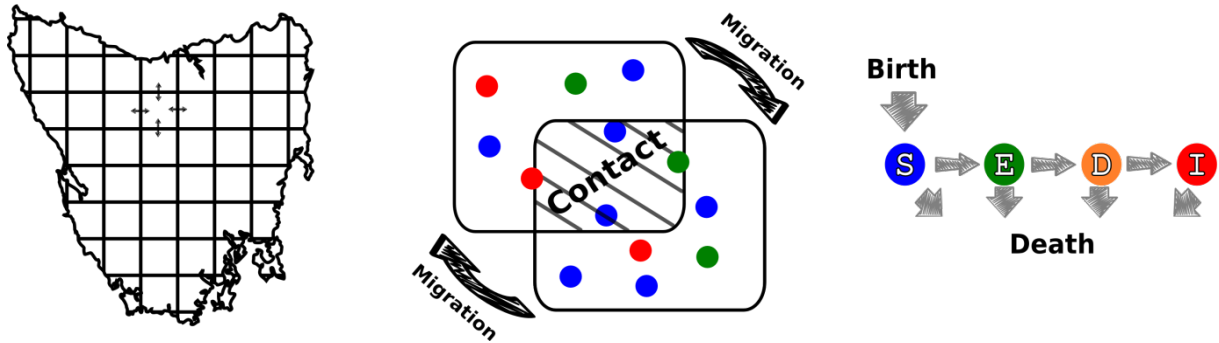

**Supplementary Figure S1.** Caricature of the model.

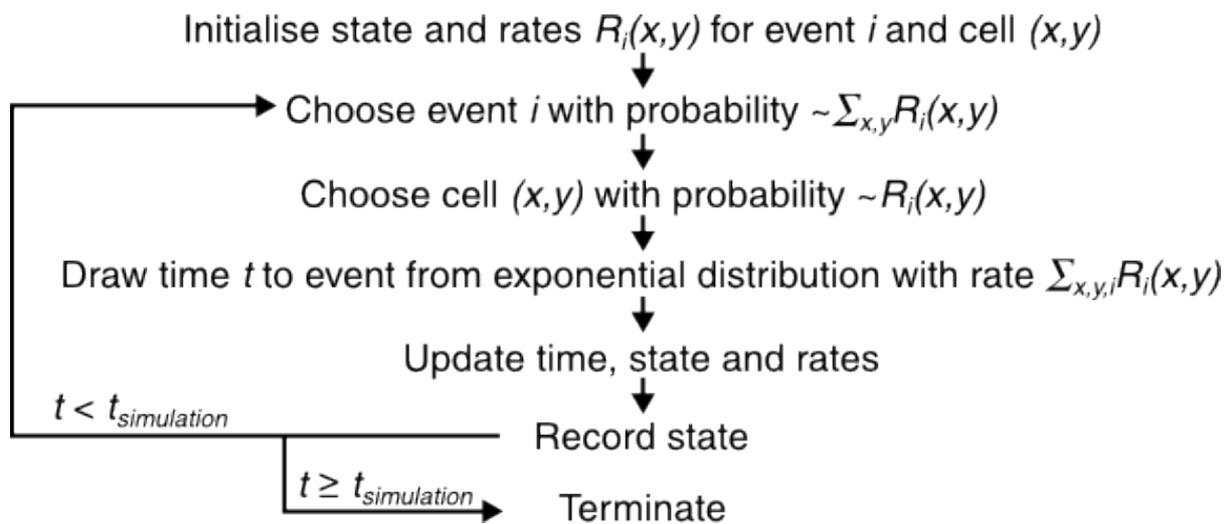

**Supplementary Figure S2.** Flowchart of the Gillespie algorithm.

**Supplementary Table S1.** Rates and effects of all events in the model.

| Event | Rate | Effect |
|---|---|---|
| Aging | $N_i(x,y)$ | $N_i(x,y) \to N_i(x,y) - 1$, $N_{i+1}(x,y) \to N_{i+1}(x,y) + 1$ |
| Natural birth | $\left(1 - \frac{\sum_i N_i(x,y)}{K(x,y)}\right) \sum_i b_i N_i(x,y)$ | $N_0(x,y) \to N_0(x,y) + 1$ |
| Natural death | $d_i N_i(x,y)$ | $N_i(x,y) \to N_i(x,y) - 1$ |

| Infection | $\dfrac{S_i(x,y)}{N_i(x,y)} \sum_j \eta_{ij} \left( I_j(x,y) + c \sum_{neighbours} I_j(x',y') \right)$ | $S_i(x,y) \to S_i(x,y) - 1,$ $E_i(x,y) \to E_i(x,y) + 1$ |
|---|---|---|
| Becoming diagnosable | $\dfrac{1}{\tau_E} E_i(x,y)$ | $E_i(x,y) \to E_i(x,y) - 1,$ $D_i(x,y) \to D_i(x,y) + 1$ |
| Becoming infectious | $\dfrac{1}{\tau_D} D_i(x,y)$ | $D_i(x,y) \to D_i(x,y) - 1,$ $I_i(x,y) \to I_i(x,y) + 1$ |
| Death from the tumour | $\dfrac{1}{\tau_I} I_i(x,y)$ | $I_i(x,y) \to I_i(x,y) - 1$ |
| Migration | $m N_i(x,y)$ | $N_i(x,y) \to N_i(x,y) - 1,$ $N_i(x',y') \to N_i(x',y') + 1$ |

**Supplementary Table S2.** Birth and death rates per age class.

|  | 0-1 year-old | 1-2 year-old | 2-3 year-old | 3-4 year-old | 4< year-old |
|---|---|---|---|---|---|
| Birth rate [1/year] | 0 | 0.13 | 1.3 | 1.65 | 1.21 |
| Death rate [1/year] | 0.1925703 | 0.1904623 | 0.1610143 | 0.1491002 | 0.3277331 |

**Supplementary Table 3.** Limits of the uniform prior for each numerical input parameter.

| Parameter | Minimum | Maximum |
|---|---|---|
| Infection rate | 0 | 50 |
| Diagnosable period | 0 | 1.2 |
| Latent period | 0 | 1 |
| Proportion infectious | 0 | 1 |
| Contact rate | 0 | 1 |
| Migration rate | 0 | 2 |

## Speed of the spatial spread

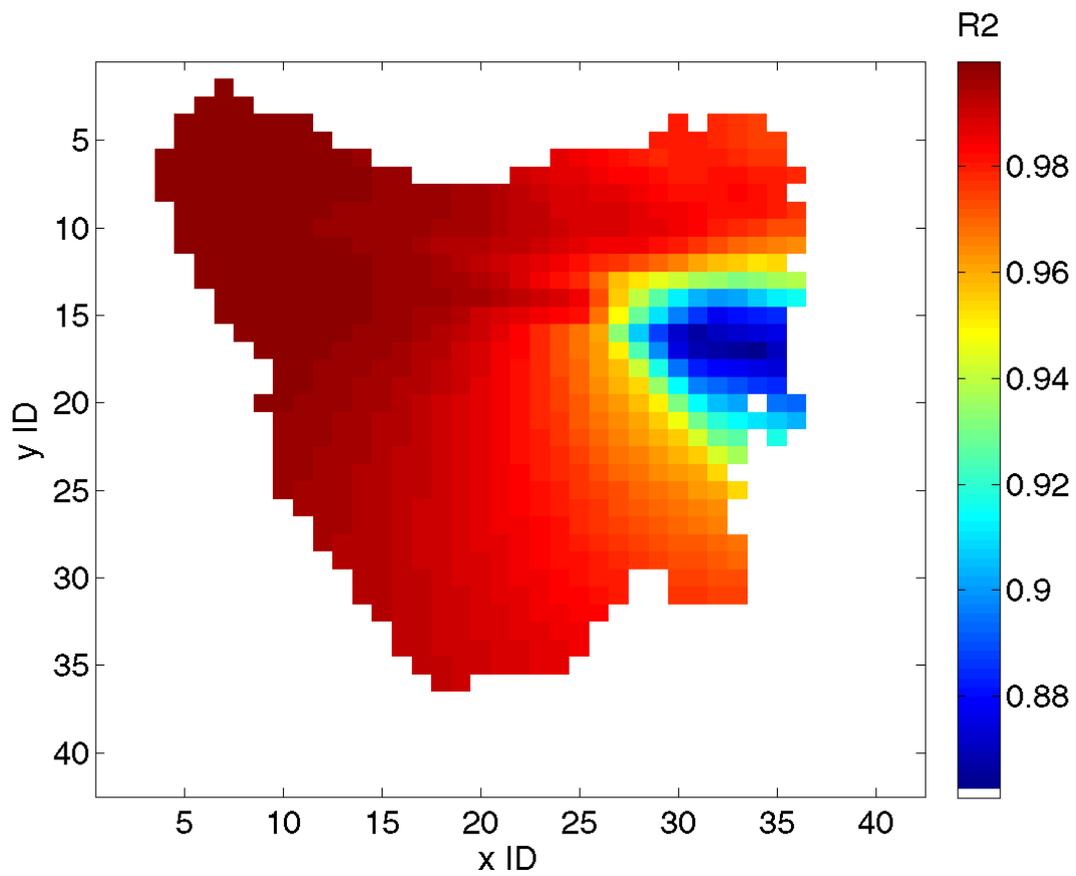

**Supplementary Figure S3.** Proportion of variance explained in the linear fit to the distance to the furthest report as a function of time, for each spatial origin. The time of the initial mutation with the highest R2 was chosen for each spatial origin.

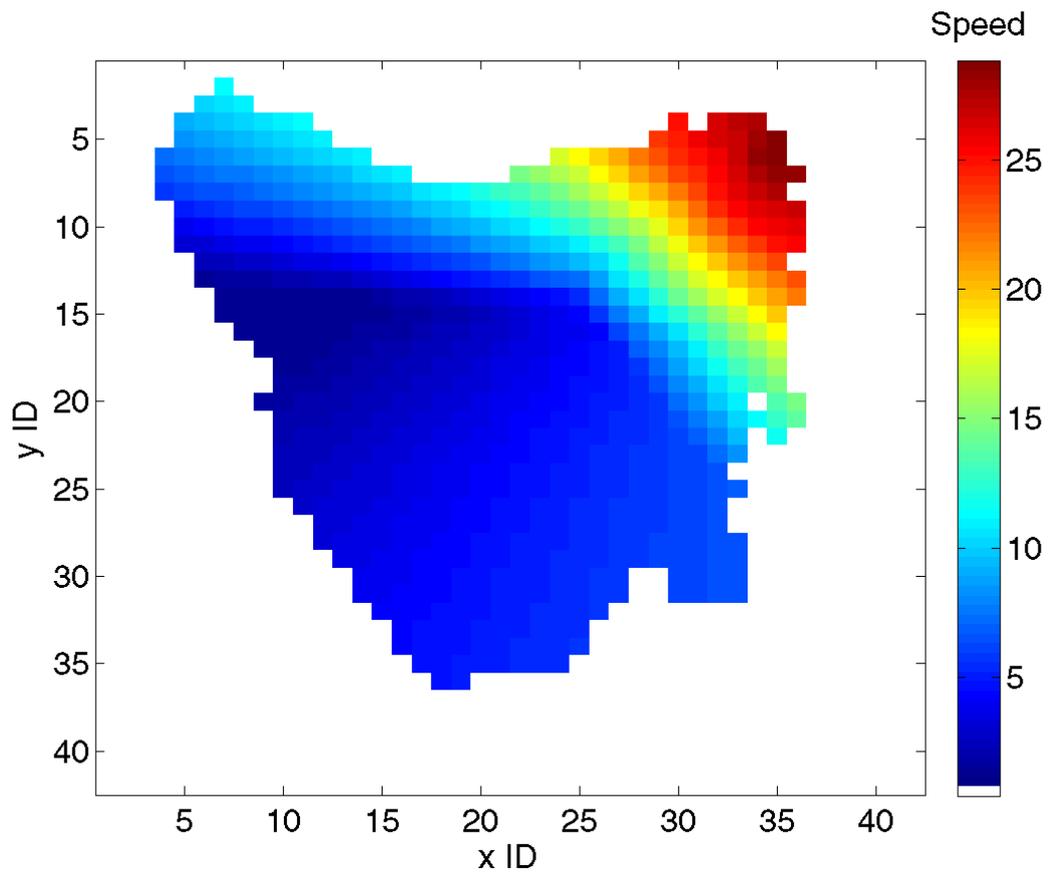

**Supplementary Figure S4.** Speed of the spatial spread, calculated from the linear fit to the distance to the furthest report as a function of time. The time of the initial mutation with the highest R2 was chosen for each spatial origin.

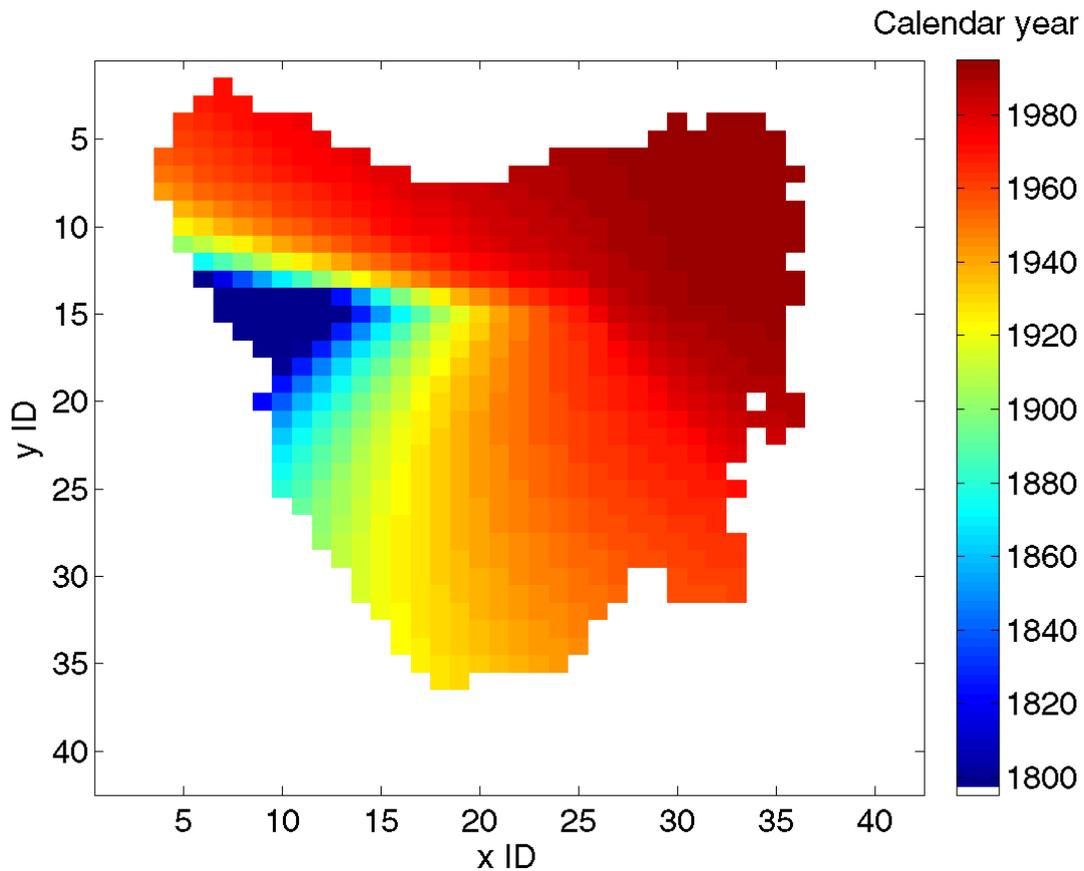

**Supplementary Figure S5.** Time of the initial mutation for each spatial origin, calculated from the linear fit to the distance to the furthest report as a function of time. The time of the initial mutation with the highest R2 was chosen for each spatial origin.

## Fitting the model

**Supplementary Table S3.** Parameters and simulated summary statistics from the 200 best-fitting parameter sets, used to generate medium- and long-term predictions. Distance: normalised distance from observed summary statistics ($r_0$=2.2644, $p_{eq}$=0.6424, $v_{data}$- $v_{simu}$=0). η: infection rate, $τ_E$: latent period, c: contact rate, m: migration rate, $τ_D$: diagnosable period, $τ_I$: infectious period, xInit: x ID of initial cell, yInit: y ID of initial cell, $r_0$: initial increase rate of adult prevalence, $p_{eq}$: equilibrial prevalence, $v_{data}$: speed of the spatial spread based on field data, assuming original mutation in cell determined by xInit and yInit, $v_{simu}$: speed of the spatial spread in the simulation.

| Distance | η [1/y] | $τ_E$ | c | m | $τ_I/(τ_D+τ_I)$ | $τ_D+τ_I$ | xInit | yInit | $r_0$ | $p_{eq}$ | $v_{data}$- $v_{simu}$ |
|---|---|---|---|---|---|---|---|---|---|---|---|
| 6.74E-05 | 45.9873 | 0.829 | 0.1299 | 1.0758 | 0.4354 | 1.0891 | 13 | 35 | 2.2439 | 0.6431 | -0.0531 |
| 0.001656 | 16.0576 | 0.5394 | 0.0727 | 0.6634 | 0.0283 | 0.8913 | 7 | 26 | 2.3556 | 0.6396 | -0.5737 |
| 0.001988 | 38.0776 | 0.8989 | 0.1716 | 0.6866 | 0.5934 | 1.1021 | 14 | 32 | 2.1944 | 0.6346 | 0.7683 |
| 0.002092 | 6.4478 | 0.4878 | 0.4028 | 0.5869 | 0.4025 | 1.0392 | 15 | 29 | 2.2945 | 0.6432 | -1.2572 |
| 0.00236 | 5.5142 | 0.6029 | 0.8302 | 0.457 | 0.2767 | 1.0633 | 16 | 35 | 2.1701 | 0.6533 | -0.0285 |
| 0.002787 | 21.8499 | 0.0506 | 0.2973 | 1.7541 | 0.8746 | 0.891 | 13 | 35 | 2.1821 | 0.6427 | 1.2152 |
| 0.003376 | 34.0109 | 0.8264 | 0.6619 | 0.7905 | 0.7279 | 1.1492 | 12 | 34 | 2.3618 | 0.6307 | 0.7859 |
| 0.00406 | 42.382 | 0.115 | 0.234 | 1.5018 | 0.9022 | 0.6722 | 6 | 30 | 2.191 | 0.6546 | -1.2442 |
| 0.004262 | 10.5561 | 0.7357 | 0.8915 | 1.7741 | 0.6394 | 0.9891 | 8 | 30 | 2.307 | 0.6228 | -0.5955 |
| 0.004577 | 32.5308 | 0.6167 | 0.5702 | 0.0333 | 0.7126 | 0.6899 | 12 | 31 | 2.0966 | 0.6352 | 0.3083 |
| 0.004676 | 48.4901 | 0.7187 | 0.089 | 0.3432 | 0.6523 | 1.1976 | 14 | 29 | 2.3135 | 0.6586 | -1.2383 |
| 0.005475 | 2.6679 | 0.4673 | 0.6625 | 1.341 | 0.0591 | 1.0352 | 13 | 34 | 2.4192 | 0.6574 | 0.0262 |
| 0.005996 | 26.3926 | 0.7688 | 0.349 | 0.9754 | 0.6124 | 1.0773 | 9 | 30 | 2.1202 | 0.6244 | 0.201801 |
| 0.0066 | 47.4632 | 0.2243 | 0.2894 | 0.2114 | 0.7641 | 0.6777 | 13 | 34 | 2.3112 | 0.6558 | -1.9321 |
| 0.006743 | 26.9415 | 0.4491 | 0.7545 | 0.1412 | 0.8003 | 0.5208 | 6 | 30 | 2.4077 | 0.6253 | 0.9541 |
| 0.007424 | 36.1149 | 0.712 | 0.193 | 0.8576 | 0.3252 | 0.7289 | 11 | 34 | 2.3772 | 0.6178 | 0.085301 |
| 0.0076 | 26.5412 | 0.771 | 0.4533 | 1.3458 | 0.7712 | 1.1862 | 7 | 27 | 2.2231 | 0.6479 | -2.3905 |
| 0.008045 | 19.8822 | 0.5389 | 0.245 | 1.7612 | 0.3437 | 0.6232 | 8 | 34 | 2.1403 | 0.6177 | 0.407599 |
| 0.00814 | 38.7164 | 0.2436 | 0.4221 | 1.2273 | 0.8595 | 0.4974 | 5 | 33 | 2.1363 | 0.6538 | -1.9304 |
| 0.008746 | 21.7242 | 0.7578 | 0.0347 | 0.1213 | 0.0923 | 1.0908 | 7 | 14 | 2.3442 | 0.6136 | -0.3438 |
| 0.009076 | 13.9525 | 0.3196 | 0.085 | 0.7304 | 0.6386 | 1.0897 | 11 | 25 | 2.0127 | 0.6457 | 0.1401 |
| 0.009377 | 18.3295 | 0.9637 | 0.6667 | 0.4578 | 0.2283 | 1.1642 | 12 | 33 | 2.2658 | 0.6125 | -0.9328 |
| 0.009918 | 15.2486 | 0.1084 | 0.068 | 0.5679 | 0.6875 | 0.8738 | 9 | 20 | 2.34 | 0.6699 | 1.2991 |
| 0.010013 | 6.7715 | 0.419 | 0.9892 | 1.8338 | 0.3529 | 0.7134 | 6 | 32 | 2.1895 | 0.6191 | -1.8373 |
| 0.010355 | 38.3518 | 0.6566 | 0.732 | 0.9451 | 0.8308 | 1.1846 | 10 | 32 | 2.211 | 0.6741 | -0.7098 |
| 0.010875 | 46.9922 | 0.9389 | 0.8191 | 0.1783 | 0.776 | 1.1206 | 10 | 33 | 2.2986 | 0.6121 | 1.3289 |
| 0.010959 | 33.8861 | 0.8765 | 0.5945 | 0.8674 | 0.4515 | 1.1365 | 8 | 33 | 2.1503 | 0.6184 | -1.7452 |
| 0.010988 | 31.8496 | 0.9424 | 0.8903 | 0.5003 | 0.6031 | 1.0593 | 13 | 35 | 2.108 | 0.6261 | -2.0224 |
| 0.011079 | 48.3174 | 0.63 | 0.0613 | 0.0028 | 0.3552 | 0.8318 | 11 | 28 | 2.0262 | 0.6387 | 1.5395 |
| 0.012032 | 23.9222 | 0.7453 | 0.0245 | 1.6705 | 0.1665 | 1.0084 | 12 | 33 | 2.0612 | 0.6607 | -1.5793 |
| 0.012102 | 41.0499 | 0.4593 | 0.183 | 0.0197 | 0.7144 | 0.7751 | 8 | 25 | 2.5258 | 0.6336 | 1.1817 |
| 0.012128 | 24.3528 | 0.8361 | 0.4367 | 0.6187 | 0.3689 | 1.0013 | 10 | 30 | 2.1629 | 0.6255 | -2.5449 |
| 0.012144 | 42.6108 | 0.744 | 0.9144 | 0.1013 | 0.9047 | 0.9391 | 10 | 28 | 2.3769 | 0.6099 | -0.6671 |
| 0.012253 | 36.203 | 0.3994 | 0.0879 | 0.8176 | 0.6334 | 0.6773 | 9 | 25 | 2.4674 | 0.6344 | -2.1708 |
| 0.012258 | 5.4226 | 0.5191 | 0.5737 | 1.5126 | 0.3071 | 0.8707 | 16 | 34 | 2.1453 | 0.6589 | -2.4977 |
| 0.01232 | 34.3651 | 0.6709 | 0.9599 | 0.9676 | 0.9343 | 1.1259 | 16 | 33 | 2.0076 | 0.632 | -1.2706 |
| 0.012975 | 15.5789 | 0.2979 | 0.3063 | 1.5988 | 0.8138 | 1.0169 | 15 | 31 | 2.3875 | 0.6207 | -2.2842 |
| 0.013371 | 47.9338 | 0.7569 | 0.1697 | 0.7067 | 0.5074 | 1.1662 | 15 | 34 | 2.4264 | 0.6486 | -2.7411 |
| 0.013403 | 23.5382 | 0.5998 | 0.2359 | 1.409 | 0.7221 | 1.1783 | 14 | 34 | 2.0448 | 0.6651 | 1.2078 |
| 0.013453 | 32.6732 | 0.5674 | 0.2835 | 0.4457 | 0.6477 | 0.8454 | 11 | 31 | 2.4917 | 0.6661 | 0.8712 |
| 0.013594 | 14.8231 | 0.2297 | 0.342 | 1.7097 | 0.753 | 0.6783 | 10 | 31 | 2.1491 | 0.6199 | 2.3801 |
| 0.013835 | 22.4686 | 0.481 | 0.1221 | 0.1531 | 0.2974 | 0.6978 | 14 | 34 | 2.4464 | 0.6591 | 2.3025 |
| 0.013898 | 19.1636 | 0.7508 | 0.313 | 1.9068 | 0.5179 | 0.9256 | 10 | 32 | 2.1622 | 0.6103 | -1.5187 |
| 0.014012 | 27.9693 | 0.6878 | 0.506 | 0.0293 | 0.6702 | 1.0734 | 10 | 27 | 2.0316 | 0.6677 | -0.5774 |
| 0.014346 | 8.2555 | 0.5689 | 0.8 | 1.0984 | 0.5574 | 0.8211 | 12 | 33 | 2.1206 | 0.6184 | -2.2151 |
| 0.014372 | 27.1066 | 0.444 | 0.3633 | 0.6743 | 0.8348 | 0.8516 | 6 | 25 | 2.1041 | 0.6123 | 1.3718 |
| 0.014386 | 42.6467 | 0.4505 | 0.3783 | 0.1512 | 0.8641 | 0.6617 | 15 | 31 | 2.2109 | 0.6045 | -0.7228 |
| 0.014486 | 41.9573 | 0.6441 | 0.3191 | 0.9528 | 0.49 | 0.9871 | 7 | 33 | 2.0996 | 0.6759 | 0.432399 |
| 0.014716 | 22.1655 | 0.6088 | 0.7591 | 0.1039 | 0.7849 | 0.9455 | 10 | 27 | 2.3916 | 0.6581 | -2.8615 |
| 0.014843 | 19.6507 | 0.5476 | 0.9822 | 1.7515 | 0.769 | 0.7301 | 10 | 34 | 2.3134 | 0.6031 | -0.3665 |
| 0.014843 | 34.8627 | 0.8373 | 0.851 | 0.2233 | 0.5341 | 0.9291 | 10 | 32 | 1.954 | 0.6388 | -0.9471 |
| 0.014857 | 30.4464 | 0.7095 | 0.4225 | 0.011 | 0.5166 | 0.9723 | 8 | 25 | 2.0463 | 0.6628 | -1.8576 |
| 0.014947 | 13.6426 | 0.6202 | 0.3019 | 0.7352 | 0.5177 | 0.9959 | 9 | 25 | 2.2042 | 0.6278 | -3.1712 |
| 0.014986 | 36.3447 | 0.5051 | 0.3334 | 1.6497 | 0.7079 | 0.8195 | 4 | 33 | 2.2586 | 0.6822 | -0.4727 |
| 0.014996 | 35.5503 | 0.5151 | 0.1113 | 0.9725 | 0.7307 | 0.9653 | 12 | 29 | 2.5689 | 0.6516 | 0.9471 |
| 0.015136 | 40.9161 | 0.8815 | 0.0248 | 0.1408 | 0.0675 | 0.9953 | 4 | 6 | 2.2736 | 0.6216 | -2.9947 |
| 0.015217 | 17.8677 | 0.6129 | 0.4575 | 0.418 | 0.511 | 1.0571 | 8 | 25 | 2.416 | 0.6744 | -1.4106 |
| 0.015615 | 18.079 | 0.8131 | 0.8727 | 0.4372 | 0.2789 | 1.0396 | 9 | 35 | 1.9395 | 0.6368 | 0.575899 |
| 0.015867 | 12.608 | 0.8537 | 0.8919 | 0.4262 | 0.6211 | 1.1424 | 10 | 26 | 2.2522 | 0.6201 | -3.0111 |
| 0.016062 | 48.6935 | 0.9069 | 0.4702 | 0.0193 | 0.802 | 1.027 | 13 | 28 | 2.0337 | 0.62 | -1.771 |
| 0.01648 | 33.6874 | 0.6952 | 0.336 | 0.7296 | 0.6213 | 1.1037 | 15 | 35 | 2.0182 | 0.6583 | -2.1205 |
| 0.016847 | 28.0309 | 0.2828 | 0.1142 | 1.6557 | 0.4459 | 0.6668 | 9 | 32 | 2.4377 | 0.6734 | -1.7245 |
| 0.017528 | 10.1094 | 0.5405 | 0.2028 | 0.7634 | 0.3047 | 1.0299 | 17 | 34 | 2.4954 | 0.6747 | 0.496 |
| 0.017826 | 17.0866 | 0.2725 | 0.1598 | 0.188 | 0.3917 | 0.5509 | 9 | 27 | 2.063 | 0.6784 | 0.229 |
| 0.018129 | 16.2973 | 0.6298 | 0.514 | 1.5023 | 0.7813 | 1.1746 | 11 | 32 | 2.1104 | 0.6545 | 3.2916 |
| 0.018317 | 24.8881 | 0.7834 | 0.322 | 0.3569 | 0.4665 | 1.0018 | 13 | 29 | 2.1801 | 0.6276 | -3.5123 |
| 0.018354 | 33.9917 | 0.6615 | 0.0664 | 0.4471 | 0.3085 | 0.7175 | 7 | 24 | 2.5971 | 0.6595 | 0.0345 |
| 0.019006 | 6.9514 | 0.4925 | 0.6009 | 1.4002 | 0.2729 | 0.713 | 8 | 35 | 2.1555 | 0.644 | 3.7381 |
| 0.01911 | 17.0014 | 0.6223 | 0.6001 | 0.2023 | 0.7225 | 1.1239 | 11 | 27 | 2.2743 | 0.6823 | -1.867 |
| 0.019238 | 34.7064 | 0.6779 | 0.0822 | 0.6261 | 0.4124 | 1.0451 | 8 | 30 | 2.4355 | 0.6547 | 3.3255 |
| 0.019302 | 26.585 | 0.9305 | 0.6015 | 0.4141 | 0.4587 | 1.0425 | 9 | 31 | 2.2338 | 0.5975 | -0.6062 |
| 0.019373 | 13.3683 | 0.5331 | 0.8664 | 1.85 | 0.8488 | 1.1292 | 12 | 29 | 2.5654 | 0.6338 | -2.1801 |
| 0.019635 | 31.5976 | 0.4664 | 0.5718 | 1.4099 | 0.8841 | 0.8328 | 4 | 30 | 2.4467 | 0.6035 | 0.849 |
| 0.019638 | 46.234 | 0.9105 | 0.3072 | 0.0468 | 0.5543 | 1.1149 | 7 | 16 | 2.1318 | 0.6283 | -3.5156 |
| 0.01972 | 2.8305 | 0.2833 | 0.4908 | 0.0004 | 0.2827 | 1.0016 | 21 | 35 | 2.4714 | 0.6478 | 3.2887 |
| 0.019816 | 16.9877 | 0.1972 | 0.4615 | 0.4393 | 0.8135 | 0.6568 | 13 | 32 | 2.2796 | 0.6479 | -3.9681 |
| 0.020518 | 15.0645 | 0.6735 | 0.7161 | 0.6567 | 0.578 | 0.8579 | 11 | 33 | 1.9949 | 0.647 | 2.85 |
| 0.020564 | 42.6484 | 0.8135 | 0.0138 | 0.9741 | 0.4784 | 1.0998 | 8 | 18 | 2.4933 | 0.6461 | -3.244 |
| 0.021132 | 22.0595 | 0.0955 | 0.0031 | 0.5155 | 0.4614 | 0.6644 | 15 | 32 | 2.1136 | 0.6608 | 3.454 |
| 0.021353 | 29.0888 | 0.2042 | 0.7266 | 0.3951 | 0.9331 | 0.8019 | 9 | 30 | 2.2666 | 0.6789 | 2.6934 |
| 0.021769 | 7.5895 | 0.6123 | 0.9906 | 1.7119 | 0.5964 | 0.9259 | 10 | 30 | 2.0949 | 0.6114 | -2.6639 |
| 0.021831 | 31.0933 | 0.504 | 0.9303 | 0.9245 | 0.8977 | 0.8103 | 15 | 35 | 2.2757 | 0.683 | -2.2928 |
| 0.021851 | 5.7506 | 0.3839 | 0.7313 | 1.0709 | 0.6395 | 0.7843 | 8 | 21 | 2.4814 | 0.6069 | -1.6809 |
| 0.022006 | 49.1676 | 0.8083 | 0.1737 | 0.4679 | 0.6952 | 1.1279 | 20 | 35 | 2.5747 | 0.6314 | -2.4243 |
| 0.022041 | 22.2222 | 0.4136 | 0.5874 | 0.0424 | 0.5825 | 0.4412 | 8 | 29 | 2.08 | 0.6459 | -3.7182 |
| 0.022564 | 11.6533 | 0.5626 | 0.9758 | 0.188 | 0.8308 | 1.0799 | 9 | 21 | 2.4703 | 0.613 | -2.6274 |
| 0.022767 | 33.6938 | 0.3059 | 0.5706 | 1.9727 | 0.881 | 0.7458 | 4 | 32 | 2.1451 | 0.6894 | -0.4495 |
| 0.022864 | 20.155 | 0.2843 | 0.3839 | 0.0465 | 0.6544 | 0.591 | 9 | 29 | 1.9992 | 0.611 | -1.7457 |
| 0.023048 | 7.9795 | 0.5836 | 0.8977 | 0.038 | 0.4732 | 0.891 | 9 | 24 | 2.265 | 0.6052 | -2.8699 |
| 0.023077 | 8.555 | 0.1768 | 0.2577 | 0.2787 | 0.4879 | 0.802 | 3 | 6 | 2.0846 | 0.6565 | -3.668 |
| 0.023128 | 40.3368 | 0.1954 | 0.4894 | 0.4855 | 0.8502 | 0.6376 | 6 | 32 | 1.8614 | 0.6404 | -0.3431 |
| 0.023215 | 30.766 | 0.5119 | 0.7694 | 0.4224 | 0.6576 | 0.7004 | 6 | 35 | 2.6183 | 0.6351 | 2.0143 |
| 0.02374 | 47.2031 | 0.0332 | 0.6703 | 1.3486 | 0.9666 | 0.8629 | 10 | 34 | 1.9239 | 0.6305 | 2.2093 |
| 0.023798 | 16.4542 | 0.5971 | 0.4521 | 0.6433 | 0.6282 | 1.1226 | 14 | 33 | 1.9216 | 0.6701 | 0.2428 |
| 0.024056 | 19.6085 | 0.4434 | 0.7148 | 0.3124 | 0.6736 | 0.5735 | 6 | 33 | 2.3792 | 0.6648 | 3.7628 |
| 0.024351 | 38.0617 | 0.9958 | 0.4882 | 1.6321 | 0.4357 | 1.0317 | 6 | 34 | 2.0023 | 0.6052 | -1.2054 |
| 0.025087 | 38.6529 | 0.8481 | 0.6488 | 0.2201 | 0.5087 | 1.1892 | 5 | 31 | 1.9524 | 0.6674 | 2.1135 |
| 0.025093 | 6.2407 | 0.4343 | 0.5742 | 0.1708 | 0.5825 | 1.0049 | 10 | 25 | 1.8525 | 0.6366 | 0.809 |
| 0.025636 | 8.2856 | 0.4039 | 0.4603 | 0.5393 | 0.6557 | 1.0202 | 5 | 10 | 2.464 | 0.6626 | -3.6188 |
| 0.025869 | 28.2048 | 0.5387 | 0.0755 | 1.2474 | 0.7425 | 0.9869 | 15 | 32 | 1.8724 | 0.6293 | 1.4396 |
| 0.026087 | 25.094 | 0.4161 | 0.7364 | 1.5403 | 0.8546 | 0.6131 | 8 | 35 | 1.8518 | 0.6392 | 1.2516 |
| 0.026326 | 29.4323 | 0.7744 | 0.7454 | 1.2806 | 0.6105 | 1.0996 | 12 | 34 | 2.2214 | 0.654 | -4.4761 |
| 0.02635 | 34.8448 | 0.3463 | 0.0165 | 0.5009 | 0.1425 | 0.5115 | 14 | 35 | 2.0472 | 0.6706 | 3.1517 |
| 0.026785 | 12.8996 | 0.6347 | 0.5159 | 0.6128 | 0.3155 | 0.9551 | 11 | 34 | 2.084 | 0.6818 | 2.5057 |
| 0.026864 | 2.2877 | 0.4054 | 0.7911 | 1.4939 | 0.1367 | 0.9945 | 10 | 35 | 2.2866 | 0.6002 | 2.8786 |
| 0.026933 | 29.7863 | 0.7445 | 0.4515 | 0.9572 | 0.6938 | 1.1704 | 13 | 34 | 1.9585 | 0.6793 | -0.9306 |
| 0.027229 | 25.2355 | 0.2398 | 0.4015 | 1.7978 | 0.8848 | 0.8685 | 13 | 34 | 2.6888 | 0.6294 | 0.4154 |
| 0.027276 | 14.8189 | 0.269 | 0.1246 | 0.3528 | 0.0624 | 0.4699 | 14 | 32 | 2.3479 | 0.6342 | -4.5522 |
| 0.027335 | 10.1662 | 0.6258 | 0.706 | 0.8643 | 0.6942 | 0.9865 | 8 | 27 | 2.5917 | 0.6087 | 1.157 |
| 0.027474 | 41.6342 | 0.0662 | 0.1178 | 0.2745 | 0.9 | 0.7668 | 19 | 30 | 2.2578 | 0.672 | -3.951 |
| 0.02752 | 42.2475 | 0.4872 | 0.2012 | 0.0924 | 0.632 | 0.7343 | 14 | 32 | 2.7018 | 0.641 | 0.6174 |
| 0.027765 | 36.9794 | 0.9497 | 0.2873 | 1.4055 | 0.5654 | 1.1901 | 8 | 31 | 1.83 | 0.6316 | 0.134401 |
| 0.028004 | 17.9174 | 0.2374 | 0.0516 | 1.7088 | 0.5754 | 0.839 | 9 | 29 | 2.6372 | 0.6716 | 0.605401 |
| 0.028007 | 14.8385 | 0.6229 | 0.9165 | 0.4492 | 0.4513 | 0.9713 | 11 | 33 | 2.1532 | 0.6947 | -0.8374 |
| 0.028345 | 26.1483 | 0.8858 | 0.5599 | 0.5441 | 0.7257 | 1.1901 | 8 | 28 | 2.5747 | 0.6105 | 2.0676 |
| 0.028429 | 3.3287 | 0.5956 | 0.941 | 0.0196 | 0.5053 | 1.1424 | 19 | 26 | 2.4499 | 0.6227 | -4.0149 |
| 0.028496 | 34.7925 | 0.4486 | 0.1711 | 1.5917 | 0.776 | 0.894 | 8 | 26 | 2.2471 | 0.687 | -2.8398 |
| 0.028704 | 4.1042 | 0.203 | 0.7189 | 0.7061 | 0.3934 | 0.7391 | 13 | 32 | 2.3769 | 0.6942 | -1.2732 |
| 0.02899 | 44.5919 | 0.6123 | 0.1941 | 0.7131 | 0.7137 | 0.9997 | 8 | 31 | 2.312 | 0.6822 | 3.3579 |
| 0.029034 | 43.9462 | 0.7867 | 0.0852 | 0.3804 | 0.6228 | 1.1635 | 8 | 17 | 2.6961 | 0.648 | -1.3959 |
| 0.029064 | 36.997 | 0.7454 | 0.7747 | 0.0512 | 0.6796 | 1.0619 | 9 | 30 | 1.9675 | 0.6847 | -0.0008 |
| 0.029473 | 29.6092 | 0.7596 | 0.7323 | 0.401 | 0.8017 | 1.0397 | 12 | 28 | 2.2601 | 0.5978 | -2.9808 |
| 0.029486 | 32.8544 | 0.6299 | 0.2048 | 1.3014 | 0.492 | 0.8768 | 14 | 32 | 2.0441 | 0.6439 | -4.2724 |
| 0.02952 | 33.0556 | 0.622 | 0.0004 | 0.424 | 0.2927 | 0.9574 | 18 | 30 | 1.8618 | 0.6502 | -2.2111 |
| 0.029646 | 15.99 | 0.5141 | 0.2755 | 0.7795 | 0.443 | 0.751 | 10 | 28 | 2.229 | 0.6067 | -3.7736 |
| 0.029786 | 20.908 | 0.8121 | 0.5216 | 1.533 | 0.5323 | 0.9223 | 14 | 34 | 2.2859 | 0.6073 | -3.8421 |
| 0.029919 | 21.6913 | 0.9664 | 0.7461 | 1.569 | 0.4519 | 1.0226 | 6 | 35 | 2.2632 | 0.5857 | -0.2453 |
| 0.030281 | 10.8678 | 0.6782 | 0.2968 | 1.1524 | 0.2677 | 0.8397 | 12 | 31 | 1.8517 | 0.6202 | -1.1516 |
| 0.030292 | 45.3869 | 0.7403 | 0.1924 | 1.7354 | 0.7965 | 1.1344 | 11 | 35 | 2.1238 | 0.611 | 3.8487 |
| 0.030305 | 10.6132 | 0.2418 | 0.8055 | 1.8763 | 0.6418 | 0.6644 | 6 | 33 | 1.8847 | 0.6725 | -1.1098 |
| 0.030357 | 26.9212 | 0.5083 | 0.6408 | 0.6405 | 0.693 | 0.5916 | 14 | 35 | 2.3548 | 0.6601 | -4.6075 |
| 0.030995 | 42.1156 | 0.6546 | 0.1396 | 1.5423 | 0.6091 | 1.025 | 8 | 30 | 2.036 | 0.6668 | -3.8226 |
| 0.031099 | 24.8964 | 0.4618 | 0.1661 | 0.4688 | 0.5937 | 0.8005 | 19 | 34 | 2.5323 | 0.6752 | -2.9764 |
| 0.031107 | 33.3501 | 0.3077 | 0.6505 | 0.4735 | 0.9335 | 0.9266 | 16 | 32 | 1.8156 | 0.6307 | -1.0552 |
| 0.031209 | 30.3833 | 0.8412 | 0.88 | 1.3566 | 0.7596 | 1.1228 | 8 | 31 | 2.7032 | 0.6277 | -1.2707 |
| 0.031331 | 12.9853 | 0.206 | 0.8734 | 0.7565 | 0.8034 | 0.4483 | 17 | 35 | 2.1844 | 0.6195 | -4.5426 |
| 0.031804 | 15.3237 | 0.3114 | 0.239 | 0.3121 | 0.5688 | 0.7387 | 8 | 21 | 1.8792 | 0.6412 | -2.9567 |
| 0.031837 | 25.8035 | 0.5149 | 0.7044 | 1.2401 | 0.7542 | 0.9067 | 8 | 32 | 1.9574 | 0.686 | -0.8404 |
| 0.032073 | 29.91 | 0.4698 | 0.0557 | 1.6709 | 0.6499 | 0.8268 | 15 | 30 | 1.9412 | 0.6541 | -3.5992 |
| 0.032158 | 26.4253 | 0.723 | 0.4934 | 0.0548 | 0.8016 | 1.1161 | 7 | 24 | 2.4145 | 0.6034 | 3.4632 |
| 0.032171 | 32.3635 | 0.8283 | 0.5714 | 0.7942 | 0.373 | 1.0689 | 9 | 33 | 1.9368 | 0.6375 | -3.6814 |
| 0.032257 | 9.6751 | 0.0632 | 0.1328 | 0.8161 | 0.5352 | 0.6787 | 9 | 22 | 2.5829 | 0.6437 | -3.8026 |
| 0.03233 | 3.2477 | 0.1883 | 0.5722 | 0.3987 | 0.2951 | 0.7704 | 7 | 24 | 1.803 | 0.6567 | -0.535 |
| 0.032383 | 44.12 | 0.7303 | 0.0191 | 0.0793 | 0.3658 | 0.9151 | 14 | 23 | 2.0659 | 0.6332 | -4.5844 |
| 0.032431 | 35.8553 | 0.3905 | 0.2678 | 0.8527 | 0.6894 | 0.6393 | 9 | 34 | 2.2829 | 0.679 | -4.0131 |
| 0.032853 | 34.2233 | 0.827 | 0.6048 | 0.3887 | 0.8159 | 1.1013 | 10 | 26 | 2.3484 | 0.6104 | -4.248 |
| 0.032862 | 45.5575 | 0.0558 | 0.1899 | 0.2435 | 0.9052 | 0.5802 | 12 | 29 | 1.8149 | 0.6618 | -0.8148 |
| 0.033016 | 14.657 | 0.0576 | 0.1589 | 0.5708 | 0.6738 | 0.6713 | 7 | 28 | 1.8536 | 0.6544 | 2.5155 |
| 0.033167 | 39.9316 | 0.8504 | 0.5554 | 0.3046 | 0.71 | 0.8977 | 10 | 33 | 1.8947 | 0.6137 | 2.2393 |
| 0.033297 | 23.1507 | 0.2495 | 0.3927 | 0.3472 | 0.8213 | 0.7101 | 14 | 34 | 1.9405 | 0.652 | 3.7717 |
| 0.033352 | 49.4286 | 0.9202 | 0.0486 | 0.0049 | 0.4742 | 1.1304 | 25 | 25 | 2.5718 | 0.6198 | -3.5101 |
| 0.033584 | 43.7003 | 0.7263 | 0.4342 | 0.523 | 0.4478 | 0.8656 | 6 | 29 | 1.8417 | 0.6431 | -2.5943 |
| 0.033602 | 11.924 | 0.0751 | 0.0076 | 0.2928 | 0.4838 | 0.8285 | 18 | 14 | 2.4415 | 0.6845 | -3.2044 |
| 0.033627 | 29.3032 | 0.7261 | 0.6477 | 1.457 | 0.777 | 1.12 | 6 | 32 | 1.8653 | 0.6571 | 2.7133 |
| 0.033854 | 23.831 | 0.229 | 0.0493 | 0.7859 | 0.7149 | 0.8258 | 18 | 31 | 2.4574 | 0.6953 | -1.4526 |
| 0.033971 | 40.3881 | 0.5966 | 0.4999 | 1.7231 | 0.8405 | 0.7994 | 9 | 30 | 1.9073 | 0.6232 | -3.181 |
| 0.034028 | 17.4654 | 0.5039 | 0.2838 | 1.6367 | 0.48 | 0.7175 | 8 | 35 | 1.8629 | 0.6323 | 2.8847 |
| 0.034213 | 20.4074 | 0.6853 | 0.3132 | 0.6039 | 0.333 | 0.9917 | 8 | 33 | 2.2409 | 0.6576 | 5.0822 |
| 0.034399 | 38.723 | 0.7498 | 0.6418 | 0.9531 | 0.8195 | 1.1459 | 14 | 31 | 1.9356 | 0.6412 | -3.9282 |
| 0.034542 | 17.2149 | 0.5941 | 0.0571 | 0.2629 | 0.3521 | 0.9149 | 8 | 9 | 2.6531 | 0.6594 | -2.9135 |
| 0.034731 | 34.4226 | 0.7357 | 0.3866 | 1.6773 | 0.7306 | 0.9646 | 10 | 34 | 2.5573 | 0.5954 | -1.3036 |
| 0.034825 | 5.7098 | 0.5554 | 0.7299 | 1.7703 | 0.3646 | 0.9562 | 7 | 34 | 2.2703 | 0.644 | 5.3004 |
| 0.034826 | 27.7938 | 0.7562 | 0.0534 | 0.1317 | 0.4417 | 1.0779 | 20 | 35 | 2.0687 | 0.6179 | 4.3875 |
| 0.035266 | 20.5688 | 0.1663 | 0.1957 | 0.7385 | 0.7369 | 0.6543 | 9 | 27 | 2.0839 | 0.6995 | -0.5661 |
| 0.035281 | 49.3462 | 0.8685 | 0.0483 | 1.0334 | 0.1841 | 1.0758 | 11 | 31 | 1.8992 | 0.6361 | -3.602 |
| 0.03536 | 16.9304 | 0.1904 | 0.0299 | 0.1035 | 0.4717 | 0.506 | 29 | 19 | 2.5284 | 0.6023 | -2.9235 |
| 0.035395 | 10.8906 | 0.4816 | 0.2537 | 0.5088 | 0.612 | 1.0382 | 10 | 20 | 1.8009 | 0.6307 | -1.7454 |
| 0.035491 | 16.2403 | 0.3306 | 0.1757 | 0.2958 | 0.3756 | 0.5051 | 18 | 35 | 2.4923 | 0.6789 | -3.57 |
| 0.035651 | 3.908 | 0.5177 | 0.5564 | 1.7615 | 0.254 | 1.0308 | 12 | 31 | 2.5603 | 0.5943 | 1.2068 |
| 0.035742 | 22.8501 | 0.6643 | 0.735 | 1.1218 | 0.718 | 1.1999 | 5 | 33 | 2.3402 | 0.6967 | 2.4705 |
| 0.03576 | 25.8706 | 0.2004 | 0.0561 | 0.9842 | 0.7584 | 0.7657 | 10 | 22 | 2.6921 | 0.6505 | -2.7412 |
| 0.035929 | 38.8314 | 0.6406 | 0.0038 | 1.2828 | 0.4563 | 1.104 | 8 | 28 | 1.9503 | 0.688 | 1.473 |
| 0.036118 | 4.5966 | 0.2092 | 0.5467 | 1.5339 | 0.4714 | 0.8298 | 8 | 25 | 2.0898 | 0.5998 | -3.4759 |
| 0.036415 | 27.5038 | 0.2488 | 0.0368 | 0.1747 | 0.7185 | 0.6552 | 21 | 12 | 2.6324 | 0.6213 | -3.2583 |
| 0.036449 | 17.4523 | 0.4734 | 0.0136 | 1.6482 | 0.1224 | 0.6266 | 18 | 35 | 2.3868 | 0.6437 | -5.2637 |
| 0.036473 | 39.3167 | 0.256 | 0.5959 | 0.3746 | 0.8914 | 0.4666 | 14 | 33 | 1.849 | 0.6506 | -3.0432 |
| 0.036555 | 42.9517 | 0.8505 | 0.8227 | 0.2682 | 0.5826 | 1.0498 | 4 | 33 | 2.1127 | 0.7018 | 0.6665 |
| 0.036589 | 19.0779 | 0.6231 | 0.3436 | 0.9767 | 0.1597 | 0.7939 | 7 | 33 | 2.4726 | 0.6044 | -3.7114 |
| 0.036606 | 28.8438 | 0.8345 | 0.9144 | 1.1999 | 0.6196 | 0.9511 | 7 | 32 | 2.2442 | 0.5976 | -3.8034 |
| 0.036763 | 44.3956 | 0.8881 | 0.3928 | 1.3176 | 0.8254 | 0.9577 | 8 | 27 | 2.1281 | 0.5819 | 0.3598 |
| 0.036804 | 35.8309 | 0.2322 | 0.3034 | 0.9099 | 0.7569 | 0.6418 | 5 | 34 | 1.9867 | 0.6829 | 2.9368 |
| 0.036954 | 24.425 | 0.4569 | 0.0656 | 0.6828 | 0.5073 | 0.8304 | 9 | 20 | 2.3353 | 0.6795 | -4.3528 |
| 0.037089 | 37.5392 | 0.8157 | 0.6983 | 0.2062 | 0.5646 | 0.9957 | 14 | 32 | 1.9018 | 0.6479 | -3.8368 |
| 0.037143 | 9.3051 | 0.2011 | 0.4382 | 1.7232 | 0.4468 | 0.5361 | 9 | 33 | 2.3516 | 0.6893 | -3.5546 |
| 0.037235 | 23.3229 | 0.451 | 0.4565 | 1.8558 | 0.8516 | 0.8619 | 13 | 30 | 2.3222 | 0.5885 | -2.812 |
| 0.03745 | 6.3178 | 0.4843 | 0.0996 | 0.2585 | 0.1121 | 1.052 | 10 | 18 | 2.7522 | 0.6265 | -1.0954 |
| 0.037475 | 25.999 | 0.7018 | 0.2885 | 0.1232 | 0.8372 | 1.1891 | 31 | 32 | 1.9508 | 0.6139 | -3.5988 |
| 0.03752 | 11.9655 | 0.774 | 0.1405 | 0.5631 | 0.2903 | 1.0728 | 9 | 26 | 1.9604 | 0.5931 | 1.2391 |
| 0.037945 | 15.3791 | 0.5881 | 0.7272 | 0.0905 | 0.5116 | 0.6215 | 12 | 33 | 2.0554 | 0.5874 | 1.7266 |
| 0.038586 | 32.5999 | 0.7863 | 0.4028 | 0.225 | 0.1939 | 1.068 | 9 | 31 | 1.9095 | 0.6658 | -3.5613 |
| 0.038631 | 38.2366 | 0.814 | 0.8794 | 1.5325 | 0.772 | 1.0656 | 9 | 32 | 2.6873 | 0.6287 | -3.0632 |
| 0.038988 | 9.7608 | 0.3904 | 0.0805 | 1.6115 | 0.0165 | 0.7676 | 10 | 33 | 1.7444 | 0.6455 | 0.7501 |
| 0.039424 | 10.358 | 0.7432 | 0.4279 | 0.4451 | 0.3996 | 1.0154 | 15 | 35 | 2.2418 | 0.6187 | 5.25 |
| 0.039472 | 31.0359 | 0.4758 | 0.1502 | 0.3377 | 0.4509 | 0.7911 | 14 | 34 | 2.7009 | 0.67 | 2.1038 |
| 0.039676 | 30.3528 | 0.6363 | 0.7809 | 0.1001 | 0.8178 | 0.8193 | 7 | 24 | 1.7912 | 0.6467 | -2.5207 |
| 0.039822 | 8.4212 | 0.3949 | 0.393 | 0.5977 | 0.4235 | 0.732 | 8 | 30 | 1.9224 | 0.627 | 4.127 |
| 0.040029 | 9.5283 | 0.5925 | 0.3825 | 0.5411 | 0.2864 | 0.9038 | 8 | 19 | 2.5178 | 0.6655 | -4.582 |
| 0.040064 | 30.7665 | 0.3617 | 0.7011 | 1.425 | 0.8915 | 0.7255 | 6 | 29 | 2.602 | 0.5966 | -1.9039 |
| 0.04019 | 23.8739 | 0.9078 | 0.1201 | 0.7124 | 0.1696 | 1.0918 | 8 | 30 | 2.2943 | 0.5949 | 3.9289 |
| 0.040236 | 16.9434 | 0.3595 | 0.3583 | 0.3327 | 0.5376 | 0.6916 | 16 | 31 | 2.4006 | 0.6534 | -5.428 |

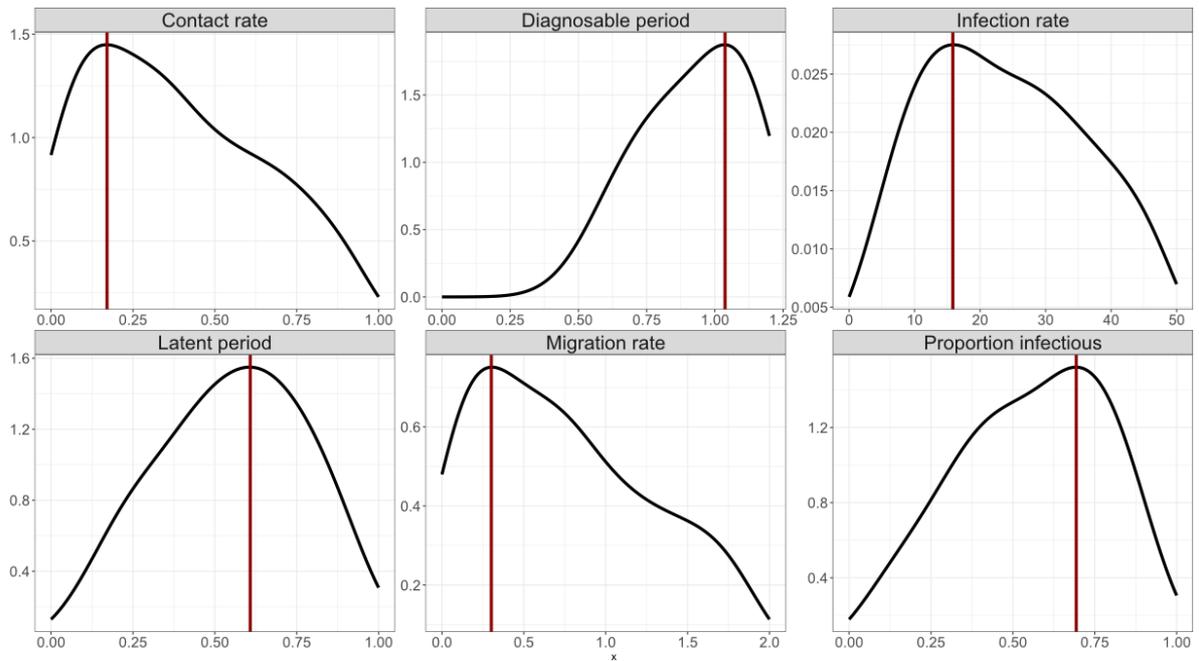

**Supplementary Figure S6.** Posterior distributions for all numerical parameters, with the median indicated in red.

**Supplementary Table S3.** Characteristics of the posterior distributions.

|  | mode | mean | median | 95% quantile lower bound | 95% quantile upper bound |
|---|---|---|---|---|---|
| **Infection rate [1/y]** | 15.829 | 0.480 | 23.151 | 2.929 | 47.105 |
| **Latent period [y]** | 0.608 | 0.547 | 0.558 | 0.100 | 0.941 |
| **Contact rate** | 0.171 | 0.411 | 0.374 | 0.023 | 0.918 |
| **Migration rate** | 0.302 | 0.403 | 0.724 | 0.044 | 1.836 |
| **Proportion infectious** | 0.693 | 0.554 | 0.571 | 0.088 | 0.942 |
| **Diagnosable period [y]** | 1.037 | 0.740 | 0.908 | 0.466 | 1.180 |

## Equilibriation of the system

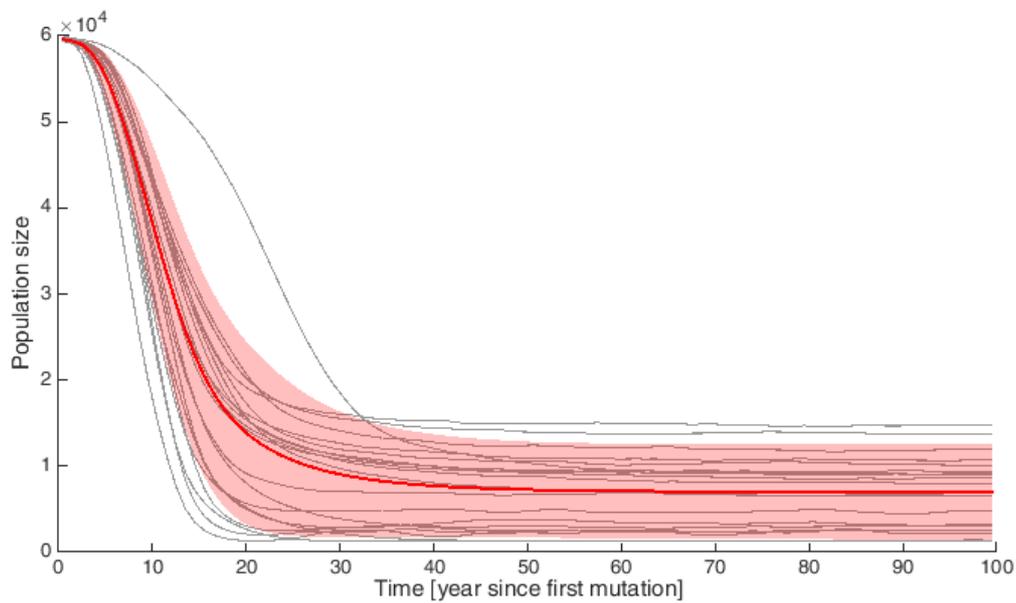

**Supplementary Figure S7.** Yearly binned time-series for total population size, from the 20 best-fitting parameter sets (grey), as well as the mean from the 200 best-fitting parameter sets used to examine the long-term behaviour (red), with the shaded area representing ±1 standard deviation.

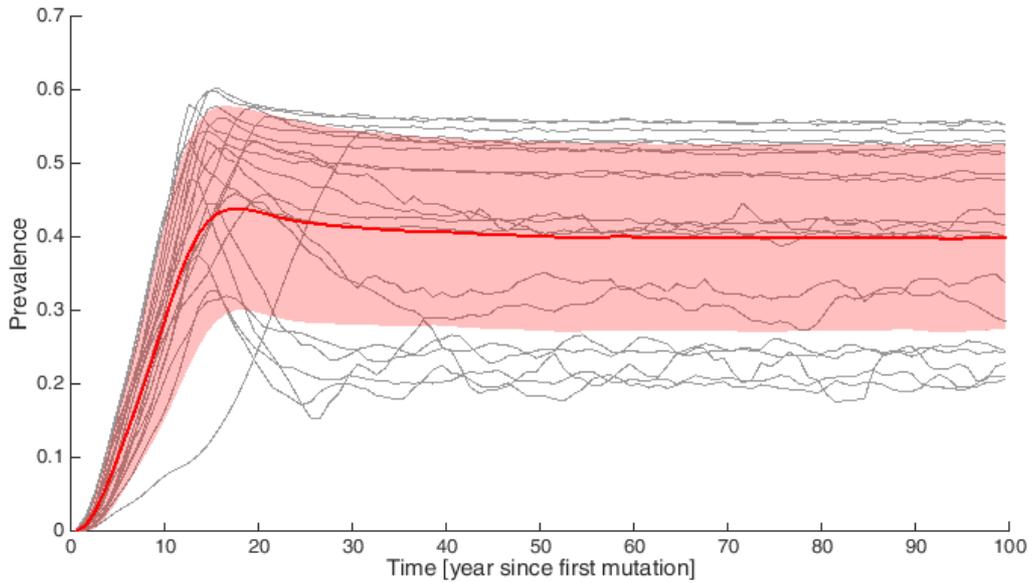

**Supplementary Figure S8.** Yearly binned time-series for total prevalence, from the 20 best-fitting parameter sets (grey), as well as the mean from the 200 best-fitting parameter sets used to examine the long-term behaviour (red), with the shaded area representing ±1 standard deviation.

## Age-structure comparison with literature

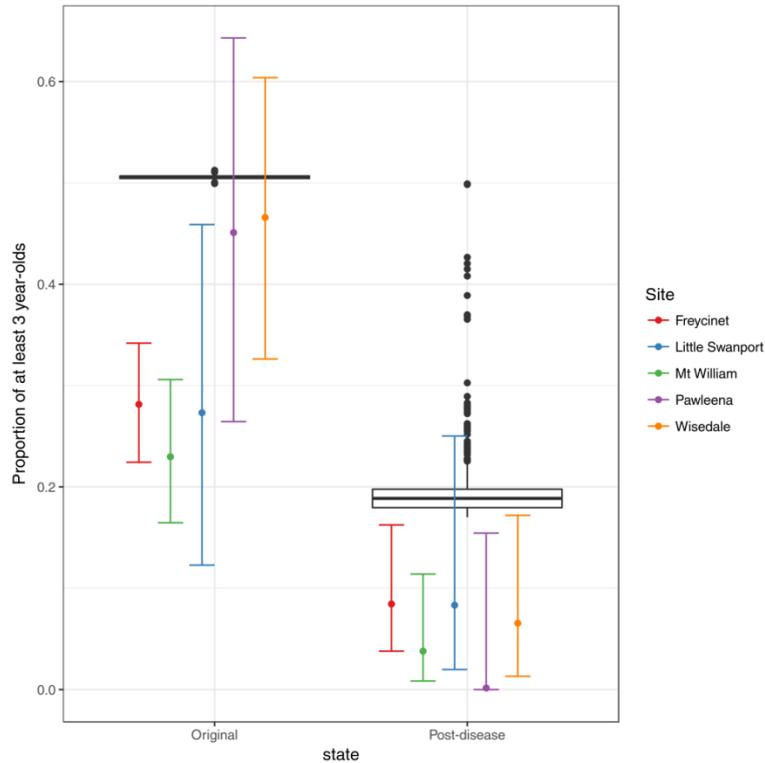

**Supplementary Figure S9.** Proportion of animals at least 3-year-old from long-term predictions of our model (black boxplot) and from field data[11].

# Determinants of long-term dynamics

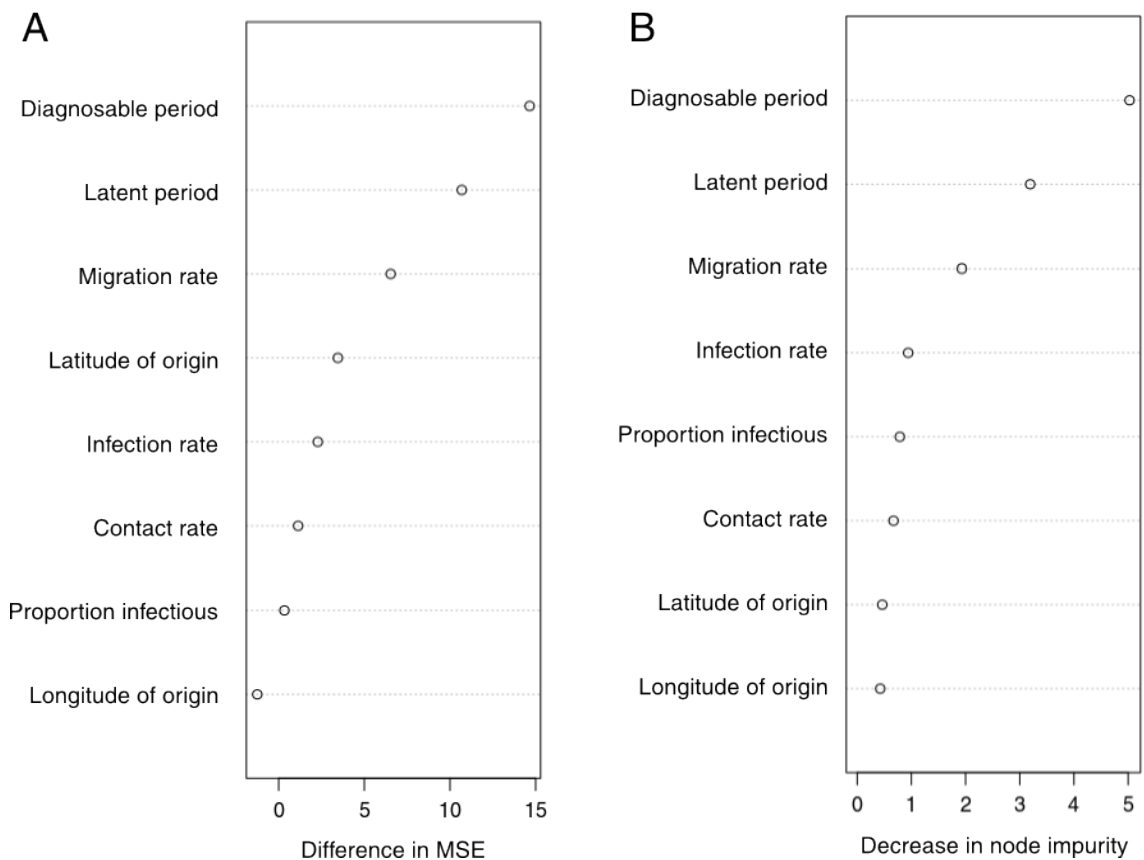

**Supplementary Figure S10. Determinants of long-term dynamics**. Measure of importance in determining cycle size for all input parameters, estimated from a forest of 200 decision trees. We used the ratio of standard deviation to mean of prevalence over the whole island as a measure of cycle size. A. Normalised difference in mean squared error after permuting the variable. B. Mean total decrease in node impurities from splitting on the variable.

# GLM for what determines cycle size

We also fitted a general linear model (GLM) to the cycle strength as a function of all input parameters using R 3.4.2[27] and calculated relative variable importance using the MuMIn 1.40.0 R package[28]. In this method, we first calculate the Akaike weights over models (in our cases, GLM-s) for every possible subsets of input parameters. Then, for each input parameter, the importance measure is the sum of Akaike weights over models including that variable. The relative importances are shown in Supplementary Table 4. The two most important variables were the total diagnosable period and the migration rate, but the latitude of origin overtook the latent period.

**Supplementary Table 4.** Importance measures in determining cycle size from the Akaike weights of GLM-s for each numerical input parameter.

|  | Total diagnosable period | Migration rate | Latitude of origin | Latent period | Infection rate | Longitude of origin | Contact rate | Proportion infectious |
| --- | --- | --- | --- | --- | --- | --- | --- | --- |
| Importance | 1.00 | 1.00 | 0.83 | 0.68 | 0.61 | 0.38 | 0.37 | 0.33 |